\documentstyle[sprocl,epsf,rotate]{article}
\def\lsim{\mathrel{\rlap{\lower4pt\hbox{\hskip1pt$\sim$}}
    \raise1pt\hbox{$<$}}}         %less than or approx. symbol
\def\gsim{\mathrel{\rlap{\lower4pt\hbox{\hskip1pt$\sim$}}
    \raise1pt\hbox{$>$}}}         %greater than or approx. symbol

\input psfig.sty

\def\Journal#1#2#3#4{{#1} {\bf #2}, #3 (#4)}

\def\NPB{{\em Nucl. Phys.} B}
\def\PLB{{\em Phys. Lett.}  B}
\def\PRL{\em Phys. Rev. Lett.}
\def\PRD{{\em Phys. Rev.} D}
\def\PRC{{\em Phys. Rev.} C}

\def\ARAA{\em Ann. Rev. Astron. Astrophys.}
\def\AJ{\em Ap. J.}
\def\AJS{\em Ap. J. Suppl.}
\def\RMP{\em Rev. Mod. Phys.}
\def\N{\em Nature}
\def\SJNP{\em Sov. J. Nucl. Phys.}

\def\AA{\em Astron. Astrophys.}
\def\AJP{\em Am. J. Phys.}
\def\JETP{\em JETP}

\def\be{\begin{equation}}
\def\ee{\end{equation}}
\def\bea{\begin{eqnarray}}
\def\eea{\end{eqnarray}}
\begin{document}

\title{Nuclear Problems in Astrophysics}

\author{W. C. Haxton}

\address{Institute for Nuclear Theory, Box 351550, and
Department of Physics, Box 351560\\
University of Washington, Seattle, WA 98195, USA\\
E-mail: haxton@phys.washington.edu}

%%%%%%%%%%%%%%%%%%%%%%%%%%%%%%%%%%%%%%%%%%%%%%%%%%%%%%%%%%%%%%
% You may repeat \author \address as often as necessary      %
%%%%%%%%%%%%%%%%%%%%%%%%%%%%%%%%%%%%%%%%%%%%%%%%%%%%%%%%%%%%%%

\maketitle\abstracts{These lectures, presented at the International
School of Physics ``Enrico Fermi,'' deal with two major themes.
The first is the remarkable story of the solar neutrino problem,
which (along with the atmospheric neutrino anomaly) recently led
to the discovery of massive neutrinos and neutrino oscillations,
physics beyond the standard model.  I will describe the physics
of the standard solar model (SSM), the experimental program that was
motivated by the discrepancies between SSM predictions
and the initial observations of Raymond Davis, Jr., and
his colleagues, and the recent results of SNO and SuperKamiokande.
These first lectures end with a description of what we have learned
about neutrino oscillations and the neutrino mass matrix,
as well as the open questions (neutrino charge conjugation
properties, the absolute mass scale, CP violation) that could 
ultimately impact our understanding of baryogenesis, the origin
of large-scale structure, and other topics in cosmology and
astrophysics.
The second theme is the core-collapse supernova 
mechanism and associated nucleosynthesis.  This problem 
connects neutrino physics, which controls much of the nuclear
physics of the star, with the long-term chemical evolution 
of our galaxy.  In particular, the $r$-process, which produces 
about half of the heavy elements, remains poorly understood, despite
important new constraints from studies of metal-poor halo stars.
The possible role of new neutrino properties on both the 
explosion mechanism and nucleosynthesis is noted.}

\section{Introduction}

These lectures were presented at the International School of
Physics ``Enrico Fermi,'' August 6-16, 2002, 
``From Nuclei and their Constituents to Stars.''  This written
version includes a few new results from later in 2002, 
such as the recent announcement by the KamLAND collaboration.

The main theme of these lectures is the interplay between neutrino
properties  --- their mass, mixing, and behavior under charge
conjugation and CP --- and astrophysical phenomena.  The first
topic is the solar neutrino problem, in which the discrepancy 
between the predictions of the standard solar model (SSM) and the
results of the chlorine experiment ultimately led to the discovery
of neutrino oscillations by the SuperKamiokande and SNO 
collaborations.  
The lectures include a discussion of the basic physics of the 
SSM, the experiments on solar and atmospheric neutrinos, the 
effects of matter on neutrino oscillations, and the current status 
of our efforts to determine the neutrino mass matrix.

The second topic is one reminiscent of the solar neutrino problem
in the 1970s, the core-collapse supernova mechanism.  The
failure thusfar to develop a robust theory of supernova explosions
may indicate some basic inadequacy in our treatment of the nuclear
astrophysics (such as our inability to model the hydrodynamics
and transport realistically in three dimensions).  But surprises
could arise from new neutrino physics, as the environment (intense
neutrino fluxes, high-velocity matter flow) is significantly
different from any other we have probed.  The supernova mechanism
is important to basic astrophysics, as it controls much of the
long-term chemical evolution of our galaxy: supernovae both
synthesize and eject new elements.  Futhermore, as a source of
measureable fluxes of neutrinos of all flavors, supernovae offer
experimentalists new opportunities to test neutrino properties.
The lectures deal with supernova neutrino physics and with
the nucleosynthesis associated directly (the $\nu$-process) and
indirectly (the $r$-process) with neutrinos.

The ``live'' audience for these lectures was advanced graduate
students and postdocs in nuclear and particle physics: the material is covered at
this level and at a depth appropriate to a survey.

\section{Solar Neutrinos~\protect\cite{haxtonsn}}

The neutrino has been with us since Wolfgang Pauli's proposal,
in December, 1930, that the emission of an unobserved spin-1/2
neutral particle might explain the apparent lack of energy
conservation in nuclear beta decay
\begin{equation}
(A,Z) \rightarrow (A,Z-1) + e^+ + \nu.
\end{equation}
Enrico Fermi was present at a number of Pauli's presentations
and discussed the neutrino with him on these occasions.  In
1934, following closely Chadwick's discovery of the neutron,
Fermi proposed a theory of beta decay based on Dirac's description
of electromagnetic interactions, but with weak currents interacting
at a point, rather than at long distance through the electromagnetic
field.  Beta decay was descibed as a proton decaying to a neutron,
a phenomenon energetically possible because of nuclear binding energies,
with the emission of a positron and neutrino.  Apart from the 
absence of parity violation, which awaited discovery until 1957,
Fermi's description is a correct low-energy approximation to 
our current standard model of weak interactions.

The neutrino was connected early on to astrophysics.  As Bethe,
Critchfield, and others unraveled the stellar processes for
hydrogen burning (the pp chain and CNO cycles), stars were
recognized to be copious sources of neutrinos
\begin{equation}
4\mathrm{p} \rightarrow {}^4\mathrm{He}+ 2e^+ + 2 \nu_e.
\end{equation}
The detectability of neutrinos -- of which Pauli had
apologetically dispaired -- was established by Cowan and Reines
in 1956, and the existence of more than one flavor of neutrino
by Lederman, Schwarz, and Steinberger in 1962.  Thus, with the
development of the Glashow-Weinberg-Salam electroweak standard
model and its prediction of weak neutral currents, it was 
recognized that supernovae would emit prodigous numbers of
neutrinos of all flavors: the hot protoneutron star at the core 
of the collapse contains a thermal sea of trapped neutrinos
produced via neutal currents.
Questions of astrophysics and of neutrino properties are thus 
interwined, and much can be learned about one by 
improving our understanding of the other.

More than three decades ago Ray Davis, Jr. and his
collaborators~\cite{davis} constructed a 0.615 kiloton C$_2$Cl$_4$
radiochemical solar neutrino detector in the Homestake Gold Mine, one
mile beneath Lead, South Dakota.  This experiment began a field,
neutrino astrophysics, that in the past few years has produced
remarkable discoveries showing that our current standard 
electroweak model is incomplete.  The new neutrino properties
that have been established -- neutrinos are massive and neutrino
flavors strongly mix -- provide our first glimpse of new interactions
that likely reside at energies far beyond the limits of current
accelerators.  Furthermore, the discoveries have important 
implications for particle dark matter, large-scale structure, 
and models where the baryon number asymmetry is connected to
leptogenesis.

The first few years of the Davis experiment showed 
that the number of neutrinos detected was considerably below the
predictions of the SSM, that is, the standard theory
of main sequence stellar evolution.
Those results were refined over the 30-year lifetime of the 
experiment, which ultimately achieved a precision equivalent to 
about 3\% of the SSM prediction.  The Cl experiment was followed  
by the SAGE~\cite{sage} and
GALLEX/GNO~\cite{gallex,GNO} gallium experiments, the Kamiokande~\cite{k}
and SuperKamiokande~\cite{sk} water Cerenkov detectors, and the SNO
heavy water Cerenkov detector~\cite{SNO}.  Furthermore the explanation for
the discrepancy that was inexorably forced on us -- massive neutrinos
oscillating as they travel from the sun's core to the earth --
proved to account for a similar anomaly in atmospheric neutrino
measurements~\cite{skatm}.  These new neutrino properties are now being 
confirmed with accelerator and reactor neutrino experiments,
such as K2K and KamLAND.  Ultimately a series of long- and 
short-baseline accelerator/reactor experiments coupled with new astrophysical measurements 
will yield precise values for the mixing parameters and probe
new phenoma, such as CP violation.

My goal in the first half of these lectures is to summarize the
solar neutrino problem -- the standard solar model, the observations,
and the accumulated evidence that led to neutrino oscillations as a
resolution.  The status of our knowledge of the neutrino mass 
matrix will be given, as well as the exciting set of outstanding
questions -- particle-antiparticle conjugation properties, CP
violation, absolute scale of neutrino mass, the mass hierarchy, implications for 
dark matter and large-scale structure, connections with 
baryogenesis -- remaining to be resolved.  This is one of those
wonderful times in physics where not only are new discoveries in
hand, but we know we have the capacity experimentally to seek
solutions to even deeper questions.

\subsection{The Standard Solar Model~\protect\cite{bbp98}}

Observations of stars reveal a wide variety of stellar conditions,
with luminosities relative to solar spanning a range $L \sim 10^{-4}$ to
$10^{6} L_\odot$ and surface temperatures $T_s \sim 2000-50000$K.
The simplest relation one could propose between luminosity and
$T_s$ is that for a blackbody
\begin{eqnarray}
L = 4 \pi R^2 \sigma T_s^4 \Rightarrow \nonumber \\
{L \over L_\odot} = ({R \over R_\odot})^2 ({T_s \over T_\odot})^4,
\end{eqnarray}
which suggests that stars of a similar structure might lie along
a one-parameter path (corresponding to $R/R_\odot$ above) in the 
luminosity (or magnitude) vs. temperature (or color) plane.  In fact, there is
a dominant path in the Hertzsprung-Russell color-magnitude 
diagram along which roughly 80\% of the stars lie.  This is
the main sequence, those stars supporting themselves by 
hydrogen burning through the pp chain or CNO cycles.

As one such star, the sun is an important test of our theory
of main sequence stellar evolution: its properties -- age, mass, surface
composition, luminosity, and helioseismology -- 
are by far the most accurately known among the stars.
The SSM traces the evolution of the Sun over the past 4.6 billion
years of main sequence burning, thereby predicting the present-day
temperature and composition profiles of the solar core that govern
neutrino production.  Standard solar models share four basic
assumptions:
    
\noindent
$\bullet$ The sun evolves in hydrostatic equilibrium, maintaining a local
balance between the gravitational force and the pressure gradient.  To
describe this condition in detail, one must specify the equation of
state as a function of temperature, density, and composition.

\noindent
$\bullet$ Energy is transported by radiation and convection.  While the solar
envelope is convective, radiative transport dominates in the core
region where thermonuclear reactions take place.  The opacity depends
sensitively on the solar composition, particularly the abundances of
heavier elements.

\noindent
$\bullet$ Thermonuclear reaction chains generate solar energy.  The standard
model predicts that over 98\% of this energy is produced from the pp
chain conversion of four protons into $^4$He (see fig. 1),
with proton burning through the CNO cycle contributing the remaining
2\%.  The Sun is a large but slow reactor: the core temperature, $T_c
\sim 1.5 \cdot 10^7$ K, results in typical center-of-mass energies for
reacting particles of $\sim$ 10 keV, much less than the Coulomb
barriers inhibiting charged particle nuclear reactions.  Thus reaction
cross sections are small: in most cases, as laboratory measurements
are only possible at higher energies, cross section data must be
extrapolated to the solar energies of interest.
    
\noindent
$\bullet$ The model is constrained to produce today's solar radius, mass, and
luminosity.  An important assumption of the standard model is that the
Sun was highly convective, and therefore uniform in composition, when
it first entered the main sequence.  It is furthermore assumed that
the surface abundances of metals (nuclei with A $>$ 5) were
undisturbed by the subsequent evolution, and thus provide a record of
the initial solar metallicity.  The remaining parameter is the initial
$^4$He/H ratio, which is adjusted until the model reproduces the
present solar luminosity after 4.6 billion years of evolution.  The
resulting $^4$He/H mass fraction ratio is typically 0.27 $\pm$ 0.01,
which can be compared to the big-bang value of 0.23 $\pm$ 0.01.  Note
that the Sun was formed from previously processed material.

\begin{figure}[htb]
\psfig{bbllx=0.5cm,bblly=4.0cm,bburx=18cm,bbury=18.5cm,figure=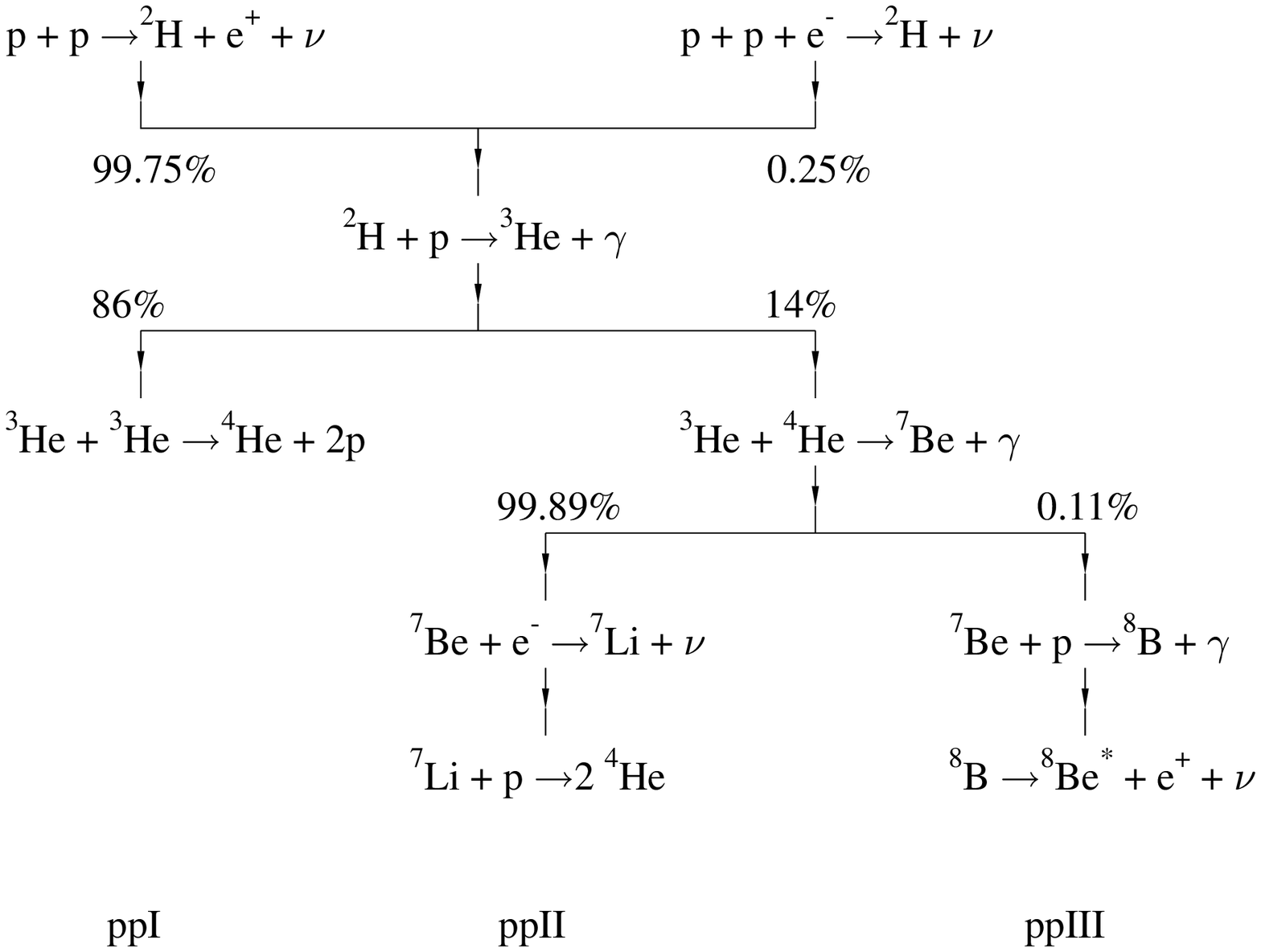,height=3.3in}
\caption{The solar pp chain.}
\end{figure}
  
The model that emerges is an evolving Sun.  As the core's chemical
composition changes, the opacity and core temperature rise, producing
a 44\% luminosity increase since the onset of the main sequence.  The
temperature rise governs the competition between the three cycles of
the pp chain: the ppI cycle dominates below about 1.6 $\cdot 10^7$ K;
the ppII cycle between (1.7-2.3) $\cdot 10^7$K; and the ppIII above
2.4 $\cdot 10^7$K.  The central core temperature of today's SSM is
about 1.55 $\cdot 10^7$K.

The competition between the cycles determines the pattern of neutrino
fluxes.  Thus one consequence of the thermal evolution of our sun is
that the $^8$B neutrino flux, the most temperature-dependent
component, proves to be of relatively recent origin: the predicted
flux increases exponentially with a doubling period of about 0.9
billion years.

A final aspect of SSM evolution is the formation of composition
gradients on nuclear burning timescales.  Clearly there is a gradual
enrichment of the solar core in $^4$He, the ashes of the pp chain.
Another element, $^3$He, is
produced and then consumed in the pp chain, eventually reaching some
equilibrium abundance.  The timescale for equilibrium to be
established as well as the eventual equilibrium abundance are both
sharply decreasing functions of temperature, and thus increasing
functions of the distance from the center of the core.  Thus a steep
$^3$He density gradient is established over time.

The SSM has had some notable successes.  From helioseismology the
sound speed profile $c(r)$ has been very accurately determined for the
outer 90\% of the Sun, and is in excellent agreement with the SSM.
Such studies verify important predictions of the SSM, such as the
depth of the convective zone.  However the SSM is not a complete model
in that it does not explain all features of solar structure, such as
the depletion of surface Li by two orders of magnitude.  This is
usually attributed to convective processes that operated at some epoch
in our sun's history, dredging Li to a depth where burning takes
place.
  
The principal neutrino-producing reactions of the pp chain and CNO
cycle are summarized in Table 1.  The first six reactions produce
$\beta$ decay neutrino spectra having allowed shapes with endpoints
given by E$_\nu^{\rm max}$.  Deviations from an allowed spectrum occur
for $^8$B neutrinos because the $^8$Be final state is a broad
resonance.  The last two reactions produce line sources of electron
capture neutrinos, with widths $\sim$ 2 keV characteristic of the
temperature of the solar core.  Measurements of the pp, $^7$Be, and
$^8$B neutrino fluxes will determine the relative contributions of the
ppI, ppII, and ppIII cycles to solar energy generation.  As discussed
above, and as later illustrations will show more clearly, the
competition between these cycles is governed in large classes of solar models by a single
parameter, the central temperature $T_c$.  The flux predictions of the
1998 calculations of Bahcall, Basu, and Pinsonneault~\cite{bbp98}
(BP98) and of Brun, Turck-Chieze and Morel~\cite{tcl} are included in
Table 1.

\begin{table}[t]
\caption{Solar neutrino sources and the flux predictions of the
Bahcall/Pinsonneault (BP98) and Brun/Turck-Chieze/Morel (BTCM98) SSMs in 
cm$^{-2}$s$^{-1}$.}
\vspace{0.2cm}
\begin{center}
\begin{tabular}{|c|c|c|c|}
\hline
 & & & \\
Source & E$_\nu^{max}$ (MeV) & BP98 & BTCM98 \\
& & & \\
\hline
& & & \\
p + p $\rightarrow ^2$H + e$^+ + \nu$ & 0.42 & 5.94E10 & 5.98E10 \\
$^{13}$N $\rightarrow ^{13}$C + e$^+ + \nu$ & 1.20 & 6.05E8 & 4.66E8 \\
$^{15}$O $\rightarrow ^{15}$N + e$^+ + \nu$ & 1.73 & 5.32E8 & 3.97E8 \\
$^{17}$F $\rightarrow ^{17}$O + e$^+ + \nu$ & 1.74 & 6.33E6 & \\
$^8$B $\rightarrow ^8$Be + e$^+ + \nu$ & $\sim$ 15 & 5.15E6 & 4.82E6 \\
$^3$He + p $\rightarrow ^4$He + e$^+ + \nu$ & 18.77 & 2.10E3 & \\
$^7$Be + e$^- \rightarrow ^7$Li + $\nu$ & 0.86 (90\%) & 4.80E9 & 4.70E9 \\
 & 0.38 (10\%) & & \\
p + e$^-$ + p $\rightarrow ^2$H + $\nu$ & 1.44 & 1.39E8 & 1.41E8 \\
 & & & \\
\hline
\end{tabular}
\end{center}
\end{table}
  
\subsection{Solar Neutrino Detection~\protect\cite{haxtontm}}

Let us start with a brief reminder about low energy neutrino-nucleus
interactions in detectors.  Consider the charged current reaction
\begin{equation}
\nu_e + (A,Z) \rightarrow e^- + (A,Z+1)
\end{equation}
Because the momentum transfer to the nucleus is very small for solar
neutrinos, it can be neglected in the weak propagator, leading to an
effective contact current-current interaction.  If we begin with the
simplest example of a semileptonic weak process, the decay of a free neutron 
n $rightarrow$ p$ + e^-+\bar{\nu}_e$, the corresponding transition amplitude
is then
\begin{equation}
S_{fi} = {G_F \over \sqrt{2}} \cos \theta_C 
\bar{u}(\mathrm{p}) \gamma_\mu (1 - g_A \gamma_5) u(\mathrm{n})
\bar{u}(e) \gamma^\mu (1 - \gamma_5) u(\nu)
\end{equation}
where $G_F$ is the weak coupling constant measured in muon decay and
$\cos \theta_c$ gives the amplitude for the weak interaction to
connect the u quark to its first-generation partner, the d quark.  The
origin of this effective amplitude is the underlying standard model
predictions for the elementary quark and lepton currents, which are
exactly left handed.  Experiment shows that the effective coupling of
the W boson to the nucleon is governed by $\gamma_\mu (1 - g_A
\gamma_5)$, as noted above, where $g_A \sim 1.26$.  The axial coupling
is thus shifted from its underlying value by the strong interactions
responsible for the binding of the quarks within the nucleon.

If an isolated nucleon were the target, one could proceed to calculate
the cross section from the effective nucleon current given above.  The
extension to nuclear systems traditionally begins with the observation
that nucleons in the nucleus are rather non-relativistic, $v/c \sim
0.1$.  The amplitude $\bar{u}$(p)$ \gamma^\mu(1-g_A \gamma_5) u$(n) can
be expanded in powers of $p/M$.  The leading vector and axial
operators are readily found to be
\begin{eqnarray}
 \gamma_0&:&~~~1 \nonumber \\
\vec{\gamma}&:&~~~\vec{p}/M \sim v/c \nonumber \\
\gamma_0 \gamma_5&:&~~~\vec{\sigma} \cdot \vec{p}/M \sim v/c 
\nonumber \\
\vec{\gamma} \gamma_5&:&~~~\vec{\sigma} 
\end{eqnarray}
Thus it is the time-like part of the vector current and the space-like
part of the axial-vector current that survive in the non-relativistic
limit.\footnote{In a nucleus these currents must be corrected for the
  presence of meson exchange contributions.  The corrections to the
  vector charge and axial three-current, which we just pointed out
  survive in the non-relativistic limit, are of order $(v/c)^2 \sim$
  1\%.  Thus the naive one-body currents are a very good approximation
  to the nuclear currents.  In contrast, exchange current corrections
  to the axial charge and vector three-current operators are of order
  $v/c$, and thus of relative order 1.  This difficulty for the vector
  three-current can be largely circumvented, because current
  conservation as embodied in the generalized Siegert's theorem allows
  one to rewrite important parts of this operator in terms of the
  vector charge operator.  In the long-wavelength limit appropriate to
  solar neutrinos, all terms unconstrained by current conservation
  vanish.  In effect, one has replaced a current operator with
  large two-body corrections by a charge operator with only small
  corrections.  In contrast, the axial charge operator is
  significantly altered by exchange currents even for long-wavelength
  processes like $\beta$ decay.  Typical axial-charge $\beta$ decay
  rates are enhanced by $\sim$ 2 because of exchange currents.}
  
If such a non-relativistic reduction is done for our single-nucleon current one
obtains
\begin{eqnarray}
S_{fi}  & \sim &  \cos \theta_c {G_F \over \sqrt{2}} 
( \phi^\dagger (\mathrm{p}) \phi(\mathrm{n}) \bar{u}(e) \gamma^0(1 -\gamma_5)u(\nu) 
\nonumber \\
  & & - \phi^\dagger (\mathrm{p}) g_A \vec{\sigma} \phi(\mathrm{n}) \cdot \bar{u}(e)
\vec{\gamma}(1-\gamma_5)u(\nu) ) 
\end{eqnarray}
where the $\phi$'s are now two-component Pauli spinors for the
nucleons.  The above result can be extended to include
$\bar{\nu}_e$ reactions by introducing the isospin operators
$\tau_\pm$ where $\tau_+$ $\mid$ n$\rangle$ = $\mid$ p$\rangle$ and
$\tau_-$$\mid$ p$\rangle$ = $\mid$ n$\rangle$, with all other matrix
elements being zero.  Thus we can generalize our n$ \rightarrow $p
amplitude to n$ \leftrightarrow $p by
\[ \phi^\dagger (\mathrm{p}) \phi(\mathrm{n}) \rightarrow \phi^\dagger (\mathrm{N}) \tau_\pm 
\phi(\mathrm{N}) \]
\[ \phi^\dagger (\mathrm{p}) \vec{\sigma} \phi(\mathrm{n}) \rightarrow \phi^\dagger (\mathrm{N})
\vec{\sigma} \tau_\pm \phi(\mathrm{N}). \] 
This result easily generalizes to nuclear decay.  Given our comments
about exchange currents, the first step is the replacement
\[ \tau_\pm \rightarrow \sum_{i=1}^A \tau_\pm(i) \]
\[ \sigma \tau_\pm \rightarrow \sum_{i=1}^A \sigma(i)
\tau_\pm(i). \] 
Plugging $S_{fi}$ into the standard cross section formula (which
involves an average over initial and sum over final nuclear spins of
the square of the transition amplitude) then yields the allowed
squared nuclear matrix element
\begin{equation}
{1 \over 2J_i+1} \left(|\langle f || \sum_{i=1}^A \tau_\pm (i) || i 
\rangle |^2
+ g_A^2 |\langle f || \sum_{i=1}^A \sigma(i) \tau_\pm(i) || i 
\rangle|^2\right).
\end{equation}
  
The Fermi operator is proportional to the isospin raising/lowering
operator: in the limit of good isospin, which typically is broken at the $\lsim$ 5\%
level for transitions between well-bound nuclear states, the Fermi operator only
connects states in the same isospin multiplet, that is, states with a
common spin-spatial structure.  If the initial state has isospin
$(T_i, M_{Ti})$, this final state has $(T_i, M_{Ti} \pm 1)$ for
$(\nu,e^-)$ and $(\bar{\nu},e^+)$ reactions, respectively, and is
called the isospin analog state (IAS).  In the limit of good isospin
the sum rule for this operator in then particularly simple
\begin{equation}
{1 \over 2J_i+1} \sum_f | \langle f || \sum_{i=1}^A \tau_+(i) || i 
\rangle |^2 =
{1 \over 2J_i+1} | \langle IAS || \sum_{i=1}^A \tau_+(i) || i 
\rangle |^2 = |N-Z|. 
\end{equation}
The excitation energy of the IAS relative to the parent ground state
can be estimated accurately from the Coulomb energy difference
\begin{equation}
E_{IAS} \sim \left({1.728 Z \over 1.12A^{1/3} + 0.78} - 1.293\right) 
\mathrm{MeV}. 
\end{equation}
The angular distribution of the outgoing electron for a pure Fermi
$(N,Z) + \nu \rightarrow (N-1,Z+1) + e^-$ transition is 1 + $\beta
\cos \theta_e$, and thus forward peaked.  Here $\beta$ is the
electron velocity.

The Gamow-Teller (GT) response is more complicated, as the operator
connects the ground state to many states in the final nucleus.  In
general we do not have a precise probe of the nuclear GT response
apart from weak interactions themselves.  However a good approximate
probe is provided by forward-angle (p,n) scattering off nuclei.
The (p,n) studies demonstrate that the GT strength tends to
concentrate in a broad resonance centered at a position $\delta =
E_{GT} - E_{IAS}$ relative to the IAS given by
\begin{equation}
 \delta \sim \left(7.0 -28.9 {N-Z \over A}\right)~\mathrm{MeV}. 
\end{equation}
Thus while the peak of the GT resonance is substantially above the IAS
for $N \sim Z$ nuclei, it drops with increasing neutron excess, with
$\delta \sim 0$ for Pb.  A typical value for the full width at half
maximum $\Gamma$ is $\sim$ 5 MeV.
The angular distribution of GT $(N,Z) + \nu_e \rightarrow (N-1,Z+1) +
e^-$ reactions is $3 - \beta \cos \theta_e$, corresponding to a
gentle peaking in the backward direction.
 
The above discussion of allowed responses can be repeated for neutral
current processes such as $(\nu,\nu')$.  The analog of the Fermi
operator contributes only to elastic processes, where the standard
model nuclear weak charge is approximately the neutron number.  As
this operator does not generate transitions, it is not yet of much
interest for solar or supernova neutrino detection, though there are
efforts to develop low-threshold detectors (e.g., cryogenic
technologies) for recording the modest nuclear recoil energies.  The
analog of the GT response involves
\begin{equation}
|M_{GT}^{fi}(\nu,\nu')|^2 = {1 \over 2J_i+1}
|\langle f || \sum_{i=1}^A \sigma(i) {\tau_3(i) \over 2} || i
\rangle |^2. 
\end{equation}
The operator appearing in this expression is familiar from magnetic
moments and magnetic transitions, where the large isovector magnetic
moment ($\mu_v \sim$ 4.706) often leads to it dominating the orbital
and isoscalar spin operators.

Finally, there is one purely leptonic reaction of great interest,
since it is the reaction exploited by Kamiokande and SuperKamiokande.
Electron neutrinos can scatter off electrons via both charged and
neutral current reactions.  The cross section calculation is
straightforward and will not be repeated here.  Two features of the
result are of importance for our later discussions, however.  Because
of the neutral current contribution, heavy-flavor $(\nu_\mu$ and
$\nu_\tau)$ also scatter off electrons, but with a cross section
reduced by about a factor of seven at low energies.  Second, for
neutrino energies well above the electron rest mass, the scattering is
sharply forward peaked.  Thus this reaction allows one to exploit the
position of the Sun in separating the solar neutrino signal from a
large but isotropic background.
  
As we mentioned earlier, the first experiment performed exploited
the reaction
\[ ^{37}\mathrm{Cl}(\nu,e^-)^{37}\mathrm{Ar}. \]
As the threshold for this reaction is 0.814 MeV, the important
neutrino sources are the $^7$Be and $^8$B fluxes.  The $^7$Be
neutrinos excite just the GT transition to the ground state, the
strength of which is known from the electron capture lifetime of
$^{37}$Ar.  The $^8$B neutrinos can excite all bound states in
$^{37}$Ar, including the dominant transition to the IAS residing at an
excitation of 4.99 MeV.  The strength of excite-state GT transitions
can be determined from the $\beta$ decay $^{37}$Ca$(\beta^+)^{37}$K,
which is the isospin mirror reaction to $^{37}$Cl$(\nu,e^-)^{37}$Ar.
The net result is that, for SSM fluxes, 78\% of the capture rate
should be due to $^8$B neutrinos, and 15\% to $^7$Be neutrinos.  The
measured capture rate~\cite{lande} 2.56 $\pm 0.16 \pm 0.16$ SNU (1 SNU
= 10$^{-36}$ capture/atom/sec) is about 1/3 the standard model value.

Similar radiochemical experiments were done by the SAGE, GALLEX, and GNO
collaborations using a different target, $^{71}$Ga.  The special
properties of this target include its low threshold and an unusually
strong transition to the ground state of $^{71}$Ge, leading to a large
pp neutrino cross section (see fig. 2).  The experimental capture
rates are $66^{+5.3+3.7}_{-5.2-3.2}$ and $74.1^{+5.4+4.0}_{-5.4-4.2}$ SNU for the SAGE
(data through December, 2001) and GALLEX/GNO (data through GNOI)
detectors, respectively.  The SSM prediction is about 130
SNU~\cite{bahcallb}.  Most important, since the pp flux is directly
constrained by the solar luminosity in all steady-state models, there
is a minimum theoretical value for the capture rate of 79 SNU, given
standard model weak interaction physics.  
Uncertainties in the $^{71}$Ga cross section due to $^7$Be neutrino
capture to two excited states of unknown strength
were greatly reduced by direct calibrations of both
detectors using $^{51}$Cr neutrino sources.

\begin{figure}[htb]
\psfig{bbllx=0.0cm,bblly=4.0cm,bburx=16cm,bbury=22.5cm,figure=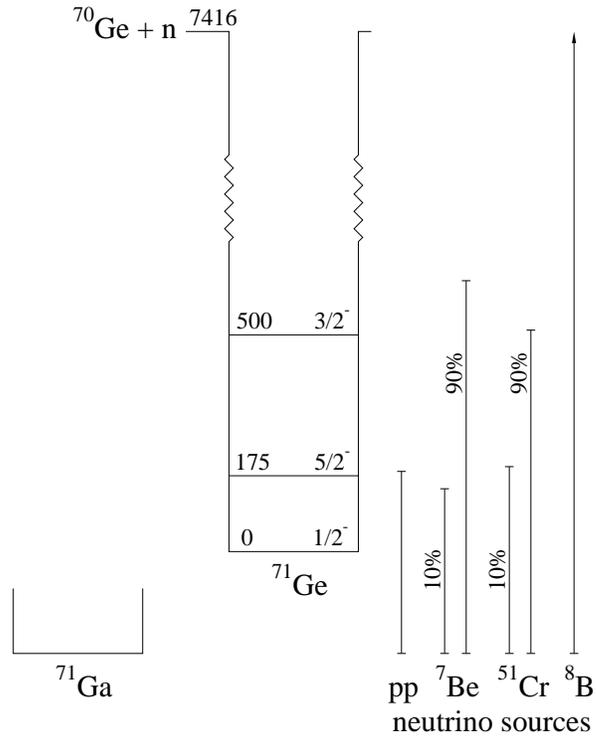,height=4.2in}
\caption{Level scheme for $^{71}$Ge showing the excited states
that contribute to absorption of pp, $^7$Be, $^{51}$Cr, and
$^8$B neutrinos.}
\end{figure}
  
The water Cerenkov experiments Kamiokande II/III and SuperKamiokande
viewed solar neutrinos on an
event-by-event basis. Solar neutrinos scatter off electrons, with the
recoiling electrons producing the Cerenkov radiation that is then
recorded in surrounding photo-tubes.  Thresholds are determined by
background rates; SuperKamiokande operated with triggers
as low as 5 MeV.  The initial experiment, Kamiokande
II/III, found a flux of $^8$B neutrinos of (2.80 $\pm 0.19 \pm 0.33)
\cdot 10^6$/cm$^2$s after about a decade of measurement.  Its much
larger successor SuperKamiokande, with a 22.5 kiloton fiducial volume,
yielded the result $(2.35 \pm 0.02 \pm 0.08) \cdot
10^6$/cm$^2$s after 1496 days of
measurements~\cite{superkamnew}, corresponding to 0.465
of the SSM $^8$B neutrino flux.
  
Results from the Sudbury Neutrino Observatory (SNO) will be
discussed later in the lectures.

\subsection{The Argument for New Particle Physics}

The pattern of solar neutrino fluxes that has emerged from these
experiments is
\begin{eqnarray}
\phi (\mathrm{pp}) & \sim & 0.9 \, \phi^{\rm {SSM}} (\mathrm{pp})\nonumber \\
\phi (^7{\rm {Be}}) & \sim & 0 \nonumber\\
\phi (^8 {\rm B}) & \sim & 0.43 \, \phi^{\rm {SSM}} (^8{\rm B}).  
\end{eqnarray}
A reduced $^8$B neutrino flux can be produced by lowering the central
temperature of the sun somewhat, as $\phi(^8$B)$\sim T_c^{18}$.
However, such an adjustment, either by varying the parameters of the
SSM or by adopting some nonstandard physics, tends to push the $\phi
(^7$Be)/$\phi(^8$B) ratio to higher values rather than the low one of
eq. (13),
\begin{equation}
{\phi (^7{\rm{Be}}) \over \phi(^8 {\rm B})} \sim T_c^{-10}.
\end{equation}
Thus the observations seem difficult to reconcile with plausible solar
model variations: one observable, $\phi(^8$B), requires a cooler core
while a second, the ratio $\phi(^7$Be)/$\phi(^8$B), requires a hotter
one.

How robust is this apparent contradiction?
In the last decade, as evidence mounted that the solar neutrino
problem was a profound one, a key issue was SSM uncertainties.
Inputs into the SSM -- pp chain nuclear
cross sections, solar parameters like the age, luminosity, and
composition, and atomic physics quantities such as opacities and
screening corrections -- all have measurement and theory errors.
It was the care with which these uncertainties were assessed that
convinced the community that the solar neutrino problem was a
serious one.
  
No issue received more scrutiny that the nuclear cross sections.
The pp chain involves a series of non-resonant charged-particle
reactions occurring at center-of-mass energies that are well below the
height of the inhibiting Coulomb barriers.  As the resulting small
cross sections generally preclude laboratory measurements at the relevant
energies, one must extrapolate higher energy measurements to threshold
to obtain solar cross sections.  This extrapolation is usually discussed
in terms of the astrophysical S-factor
\begin{equation}
\sigma (E) = {S(E) \over E} \exp (-2 \pi \eta)
\end{equation}
where $\eta = {Z_1Z_2 \alpha \over \beta}$, with $\alpha$ the fine
structure constant and $\beta = v/c$ the relative velocity of the
colliding particles.  This parameterization removes the gross Coulomb
effects associated with the s-wave interactions of charged, point-like
particles.  The remaining energy dependence of S(E) is gentle and can
be expressed as a low-order polynomial in E.  Usually the variation of
S(E) with E is taken from a direct reaction model and then used to
extrapolate higher energy measurements to threshold.  The model
accounts for finite nuclear size effects, strong interaction effects,
contributions from other partial waves, etc.  As laboratory
measurements are made with atomic nuclei while conditions in the solar
core guarantee the complete ionization of light nuclei, additional
corrections must be made to account for the different electronic
screening environments.

Thus a great deal of effort was invested in laboratory measurements,
in the theory required to extrapolate those measurements to the 
Gamow peak energies where solar reactions take place, and in
assessing the resulting cross section uncertainties.
In addition, other SSM uncertainties were evaluated.  One 
qualitative illustration of the results is given by fig. 3,
which summarizes a Monte Carlo study of SSM input parameter uncertainties.
Five key input parameters, the primordial heavy-element-to-hydrogen ratio
Z/X and S(0) for the p-p, $^3$He-$^3$He, $^3$He-$^4$He, and p-$^7$Be
were varied according to their assigned uncertainties, 
assuming for each parameter a normal distribution with the
appropriate mean and standard deviation.  (These five were the parameters assigned the
largest uncertainties.)  Smaller uncertainties from radiative
opacities, the solar luminosity, and the solar age were folded into
the results of the model calculations perturbatively~\cite{bu,bh}.

The resulting pattern of $^7$Be and $^8$B flux predictions 
produces an elongated error ellipse, showing that $\nu$ flux changes
are strongly correlated even when a large set of distinct
uncertainties are explored.  Those variations producing $\phi(^8$B) below
0.8$\phi^{\rm{SSM}}(^8$B) tend to produce a reduced $\phi(^7$Be), but
the reduction is always less than 0.8.  Thus a greatly reduced
$\phi(^7$Be) cannot be achieved within the uncertainties assigned to
parameters in the SSM.
    
\begin{figure}[htb]
\psfig{bbllx=0.5cm,bblly=4.0cm,bburx=17cm,bbury=22.5cm,figure=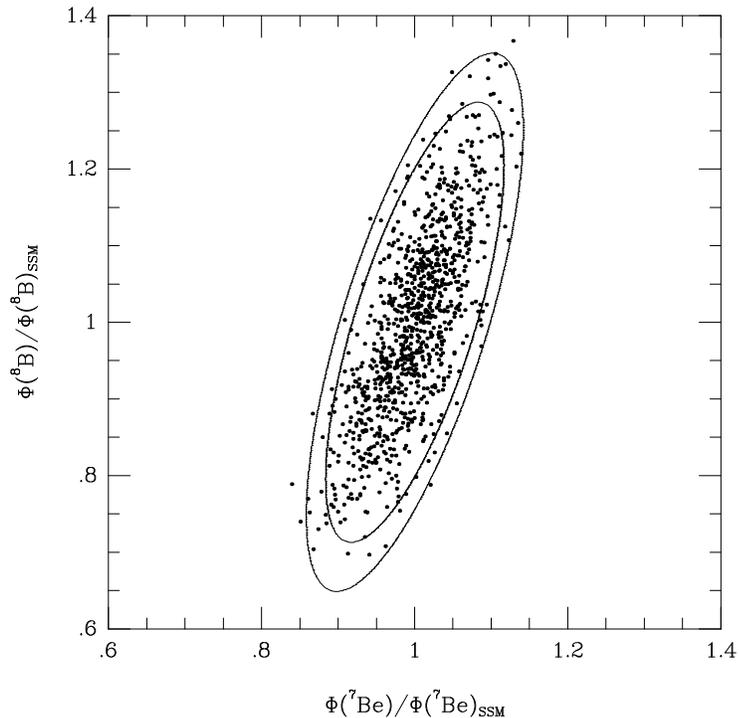,height=3.9in}
\caption{SSM $^7$Be and $^8B$ flux predictions.  The dots represent
the results of SSM calculations where the input parameters were
varied according to their assigned uncertainties, as described
in the text.  The 90\% and 99\% confidence level error ellipses
are shown.}
\end{figure}
  
A similar exploration, but including parameter variations very far
from their preferred values, was carried out by Castellani et
al.~\cite{cast}, who displayed their results as a function of the
resulting core temperature $T_c$.  The pattern that emerges is
striking (see fig. 4): parameter variations producing the same value
of $T_c$ produce remarkably similar fluxes.  Thus $T_c$ provides an
excellent one-parameter description of standard model perturbations.
Figure 4 also illustrates the difficulty of producing a low ratio of
$\phi(^7$Be)/$\phi(^8$B) when $T_c$ is reduced.

The Monte Carlo parameter variations of fig. 3 were constrained
to reproduce the solar luminosity.  Those variations show a similar
strong correlation with $T_c$
\begin{equation}
\phi(\mathrm{pp}) \propto T_c^{-1.2} ~~~~~~~  \phi(^7{\rm {Be}}) \propto T_c^8 ~~~~~~~
 \phi(^8 {\rm B}) \propto T_c^{18}.
\end{equation}
Figures 3 and 4 offer a strong argument that reasonable
variations in the parameters of the SSM, or nonstandard
changes in quantities like the metallicity, opacities, or
solar age, cannot produce the pattern of fluxes deduced
from experiment (eq. (13)).  This would seem to limit 
possible solutions to errors either in the underlying physics
of the SSM or in our understanding of neutrino properties. 

\begin{figure}[htb]
\psfig{bbllx=0.3cm,bblly=4.0cm,bburx=14.5cm,bbury=24.0cm,figure=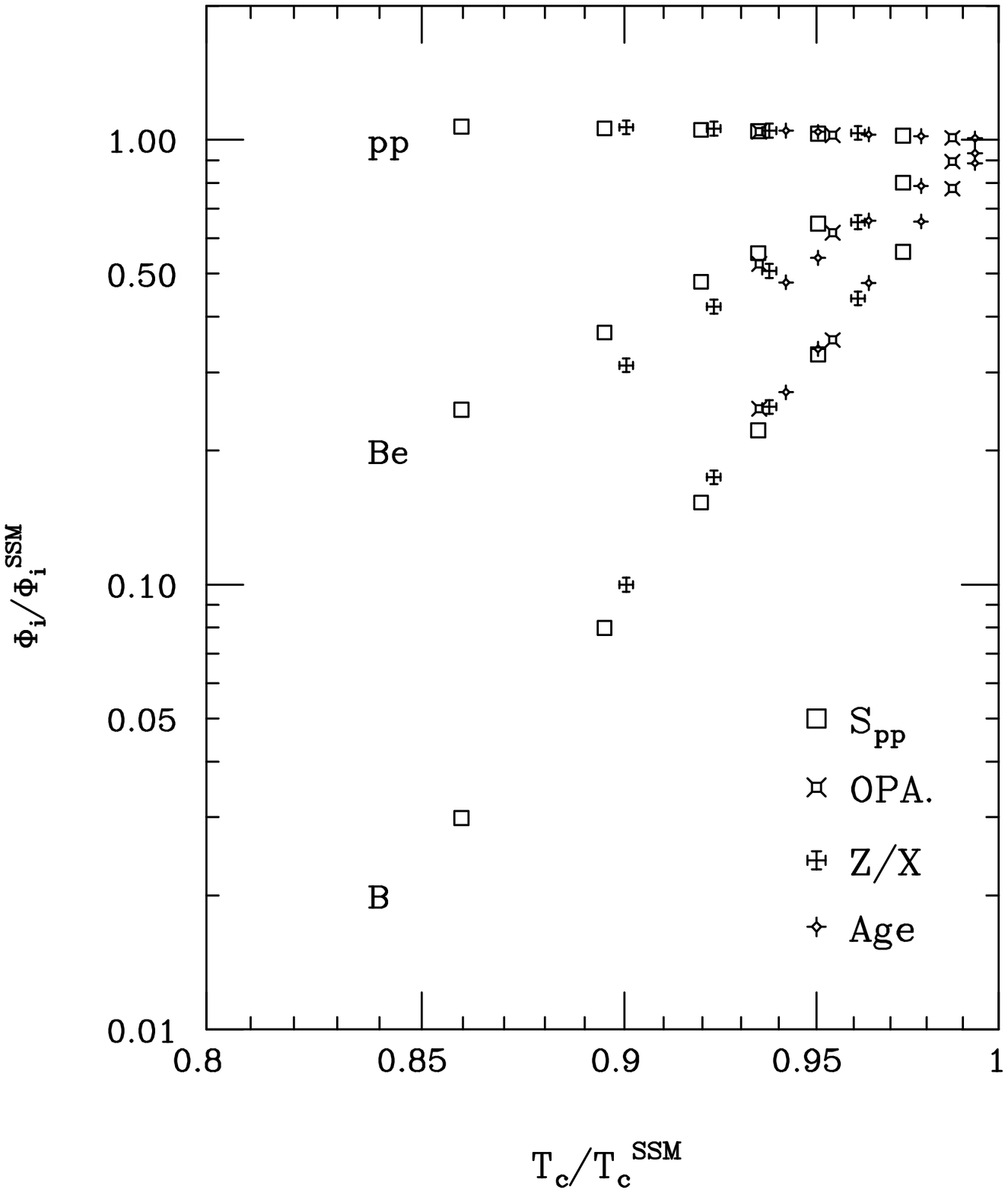,height=3.9in}
\caption{The responses of the pp, $^7$Be, and $^8$B neutrino fluxes
to the indicated variations in solar model input parameters,
displayed as a function of the resulting central temperature
$T_c$.  From Castellani et al.}
\end{figure}
  
The Castellani et al. explorations belong to a larger class of
proposed nonstandard solar models where either very large 
parameter changes (in nuclear cross sections, opacities, etc.) or 
new physics (e.g., mixing in the solar core) are hypothesized.
Once results from Cl, SAGE/GALLEX, and Kamiokande were available,
it became possible
to argue that no such nonstandard solar model can
solve the solar neutrino problem: if one assumes undistorted neutrino
spectra, no combination of pp, $^7$Be, and $^8$B neutrino fluxes fits
the experimental results well~\cite{karsten}.  In fact, in an
unconstrained fit, the required $^7$Be flux is unphysical, negative by
almost 3$\sigma$.  Thus, barring some unfortunate experimental error,
new particle physics seemed to be the indicated solution.
This conclusion was reinforced by the successes the SSM had in
reproducing the new data on helioseismology~\cite{cdhelio}.
Suggested particle physics solutions include neutrino
oscillations, neutrino decay, neutrino magnetic moments, and weakly
interacting massive particles.  Among these, the
Mikheyev-Smirnov-Wolfenstein effect --- neutrino oscillations enhanced
by matter interactions --- was widely regarded as the most plausible.

\subsection{Neutrino Oscillations}

One odd feature of particle physics is that neutrinos, which are not
required by any symmetry to be massless, nevertheless must be much
lighter than any of the other known fermions.  For instance, the
current limit on the $\overline{\nu}_e$ mass is $\lsim$ 2.2 eV.  The
standard model requires neutrinos to be massless, but the reasons are
not fundamental.  Dirac mass terms $m_D$, analogous to the mass terms
for other fermions, cannot be constructed because the model contains
no right-handed neutrino fields.  Neutrinos can also have Majorana
mass terms
\begin{equation}
\overline{\nu^c_L} m_L \nu_L ~~~ \mathrm{and} 
~~ \overline{\nu^c_R} m_R \nu_R 
\end{equation}
where the subscripts $L$ and $R$ denote left- and right-handed
projections of the neutrino field $\nu$, and the superscript $c$
denotes charge conjugation.  The first term above is constructed from
left-handed fields, but can only arise as a nonrenormalizable
effective interaction when one is constrained to generate $m_L$ with
the doublet scalar field of the standard model.  The second term is
absent from the standard model because there are no right-handed
neutrino fields.

None of these standard model arguments carries over to the more
general, unified theories that theorists believe will supplant the
standard model.  In the enlarged multiplets of extended models it is
natural to characterize the fermions of a single family, e.g.,
$\nu_e$, e, u, d, by the same (Dirac) mass scale $m_D$.  
Indeed the charged members of this multiplet all have comparable
masses $\sim$ 1 MeV.  The neutrino, however, is much lighter;
but the neutrino, which carries no charge, is the only standard
model fermion that can have both Majorana and Dirac mass
terms.  Thus it is natural to explain small neutrino masses as
a consequence of Majorana masses.
In the seesaw mechanism\cite{seesaw},
\begin{equation}
M_\nu \sim \left(\begin{array}{cc}
0 & m_D \\
m^T_D & m_R \end{array}\right) .  
\end{equation}
Diagonalization of the mass matrix produces one light neutrino,
$m_{\mathrm{light}}\sim {m_D^2 \over m_R}$, and one unobservably
heavy, $m_{\mathrm{heavy}} \sim m_R$.  The factor ($m_D$/$m_R$) is the
needed small parameter that accounts for the distinct scale of
neutrino masses.  The masses for the $\nu_e, \nu_\mu$, and $\nu_\tau$
are then related to the squares of the corresponding quark masses
$m_u$, $m_c$, and $m_t$.  Taking $m_R \sim 10^{16}$ GeV for the right-handed Majorana mass, a typical
grand unification scale for models built on groups like SO(10), the
seesaw mechanism gives the crude relation
\begin{equation}
m_{\nu_e}: m_{\nu_\mu}: m_{\nu_\tau} \leftrightarrow 2 \cdot 
10^{-12}: 2 
\cdot 10^{-7}: 3 \cdot 10^{-3} \mathrm{eV}. 
\end{equation}
The fact that solar neutrino experiments can probe small neutrino
masses, and thus provide insight into possible new mass scales $m_R$
that are far beyond the reach of direct accelerator measurements, has
been an important theme of the field.
    
One of the most interesting possibilities for solving the solar
neutrino problem has to do with neutrino masses.  For simplicity we
will discuss just two neutrinos.  If a neutrino has a mass $m$, we
mean that as it propagates through free space, its energy and momentum
are related in the usual way for this mass.  Thus if we have two
neutrinos, we can label those neutrinos according to the eigenstates
of the free Hamiltonian, that is, as mass eigenstates.

But neutrinos are produced by the weak interaction.  In this case, we
have another set of eigenstates, the flavor eigenstates.  We can
define a $\nu_e$ as the neutrino that accompanies the positron in
$\beta$ decay.  Likewise we label by $\nu_\mu$ the neutrino produced
in muon decay.

Now the question: are the eigenstates of the free Hamiltonian and of
the weak interaction Hamiltonian identical?  Most likely the answer is
no: we know this is the case with the quarks, since the different
families (the analog of the mass eigenstates) do interact through the
weak interaction.  That is, the up quark decays not only to the down
quark, but also occasionally to the strange quark.  (This is why we
had a $\cos \theta_c$ in our weak interaction amplitude: the amplitude
for $u \rightarrow s$ is proportional to $\sin \theta_c$.)  Thus we
suspect that the weak interaction and mass eigenstates, while spanning
the same two-neutrino space, are not coincident: the mass eigenstates
$|\nu_1 \rangle$ and $|\nu_2 \rangle$ (with masses $m_1$ and $m_2$)
are related to the weak interaction eigenstates by
\begin{eqnarray}
|\nu_e\rangle  &=& \cos \theta_v |\nu_1\rangle  
+ \sin \theta_v|\nu_2 \rangle  \nonumber \\
|\nu_\mu\rangle &=& - \sin \theta_v |\nu_1 \rangle 
+ \cos \theta_v |\nu_2 
\rangle 
\end{eqnarray}
where $\theta_v$ is the (vacuum) mixing angle. 
  
An immediate consequence is that a state produced as a $|\nu_e\rangle$
or a $|\nu_\mu\rangle$ at some time $t$ --- for example, a neutrino
produced in $\beta$ decay --- does not remain a pure flavor eigenstate
as it propagates away from the source.  The different
mass eigenstates comprising the neutrino will accumulate different
phases as they propagate downstream, a phenomenon known as vacuum
oscillations (vacuum because the experiment is done in free space).
To see the effect, suppose the neutrino produced in a $\beta$ decay
is a momentum eigenstate.  At time $t$=0
\begin{equation}
|\nu(t=0)\rangle  = |\nu_e \rangle = \cos \theta_v |\nu_1\rangle  
+ \sin \theta_v|\nu_2 \rangle . 
\end{equation}
Each eigenstate subsequently propagates with a phase
\begin{equation}
e^{i(\vec{k} \cdot \vec{x} - \omega t)} =
e^{i(\vec{k} \cdot \vec{x} - \sqrt{m_i^2 + k^2}t)} . 
\end{equation}
But if the neutrino mass is small compared to the neutrino
momentum/energy, one can write
\begin{equation}
\sqrt{m_i^2+k^2} \sim k(1 + {m_i^2 \over 2k^2}) . 
\end{equation}
Thus we conclude
\begin{eqnarray}
|\nu(t) \rangle &=& e^{i(\vec{k} \cdot \vec{x} - kt
-(m_1^2+m_2^2)t/4k)} \nonumber \\
& & \times [\cos \theta_v |\nu_1 \rangle e^{i \delta m^2 t/4k}
+ \sin \theta_v |\nu_2 \rangle e^{-i \delta m^2 t/4k} ] . 
\label{eq:24}
\end{eqnarray}
We see there is a common average phase (which has no physical
consequence) as well as a beat phase that depends on
\begin{equation}
\delta m^2 = m_2^2 - m_1^2 .
\end{equation}
Now it is a simple matter to calculate the probability that 
our neutrino state remains a $|\nu_e\rangle$ at time t
\begin{eqnarray}
P_{\nu_e} (t) &=& | \langle \nu_e | \nu(t) \rangle |^2 \nonumber \\ 
 &=& 1 - \sin^2 2 \theta_v \sin^2 \left({\delta m^2 t \over 4 
k}\right) \rightarrow 1 - {1 \over 2} \sin^2 2 \theta_v 
\end{eqnarray}
where the limit on the right is appropriate for large $t$.  
(When one properly describes the neutrino state as a wave packet, the
large-distance behavior follows from the eventual separation of the
mass eigenstates.)  Now $E
\sim k$, where $E$ is the neutrino energy, by our assumption that the
neutrino masses are small compared to $k$.  We can reinsert
the implicit constants to write the probability in terms of the distance $x$ of
the neutrino from its source,
\begin{equation}
P_{\nu} (x) =1 - \sin^2 2 \theta_v \sin^2 \left({\delta m^2c^4 
x\over 4 \hbar c E} \right) . 
\end{equation}
If the oscillation length
\begin{equation}
L_o = {4 \pi \hbar c E \over \delta m^2 c^4} 
\end{equation}
is comparable to or shorter than one astronomical unit, a reduction in
the solar $\nu_e$ flux would be expected in terrestrial neutrino
oscillations.
  
The suggestion that the solar neutrino problem could be explained by
neutrino oscillations was first made by Pontecorvo in 1958, who
pointed out the analogy with $K_0 \leftrightarrow \bar K_0$
oscillations.  From the point of view of particle physics, the sun is
a marvelous neutrino source.  The neutrinos travel a long distance and
have low energies ($\sim$ 1 MeV), implying a sensitivity down to
\begin{equation}
\delta m^2 \gsim 10^{-12} eV^2.
\end{equation}
In the seesaw mechanism, $\delta m^2 \sim m^2_2$, so neutrino masses
as low as $m_2 \sim 10^{-6}$ eV could be probed.  

From the expressions above one expects vacuum oscillations to affect
all neutrino species equally, if the oscillation length is small
compared to an astronomical unit.  This is somewhat in conflict with
the solar neutrino data, as we have argued that the $^7$Be neutrino
flux is quite suppressed.  Furthermore, there is a weak theoretical
prejudice that $\theta_v$ should be small, like the Cabibbo angle.
The first objection, however, can be circumvented in the case of
``just so" oscillations where the oscillation length is comparable to
one astronomical unit.  In this case the oscillation probability
becomes sharply energy dependent, and one can choose $\delta m^2$ to
preferentially suppress one component (e.g., the monochromatic $^7$Be
neutrinos), though the requirement for large mixing angles
remains.  This ``just so'' vacuum scenario is one that received
considerable attention.
  
Below we will find that oscillations in matter can lead to nearly 
total flavor conversion even if the mixing angle is small.
In preparation for this we first present
the results above in a slightly more general way.  The analog of eq.
(\ref{eq:24}) for an initial muon neutrino ($|\nu(t=0)\rangle =
|\nu_\mu\rangle$) is
\begin{eqnarray}
|\nu(t) \rangle &=& e^{i(\vec{k} \cdot \vec{x} - kt
-(m_1^2+m_2^2)t/4k)} \nonumber \\
&& \times [-\sin \theta_v |\nu_1 \rangle e^{i \delta m^2 t/4k}
+ \cos \theta_v |\nu_2 \rangle e^{-i \delta m^2 t/4k} ]
\label{eq:30}
\end{eqnarray}
Now if we compare Eqs. (\ref{eq:24}) and (\ref{eq:30}) we see that
they are special cases of a more general problem.  Suppose we write
our initial neutrino wave function as
\begin{equation}
 |\nu(t=0)\rangle = a_e(t=0) |\nu_e \rangle + a_\mu(t=0) 
|\nu_\mu \rangle . 
\label{eq:31}
\end{equation}
Then Eqs. (\ref{eq:24}) and (\ref{eq:30}) tell us that the subsequent
  propagation is described by changes in $a_e(x)$ and $a_\mu(x)$
  according to (this takes a bit of algebra)
\begin{equation}
i {d \over dx} \left( \matrix { a_{\textstyle e} \cr
a_{\textstyle \mu} \cr} \right) = {1 \over 4E} \left ( \matrix{
- \delta m^2 \cos 2 \theta_{\textstyle v}
~~~~~~~~~~\delta m^2\sin
2\theta_{\textstyle v} \cr 
\delta m^2\sin 2 \theta_{\textstyle v} ~~~~~~~~~~~ 
\delta m^2
\cos 2\theta_{\textstyle v} \cr} \right) \left( \matrix {
a_{\textstyle e} \cr
a_{\textstyle \mu} \cr} \right) . 
\end{equation}
Note that the common phase has been ignored: it can be absorbed into
the overall phase of the coefficients $a_e$ and $a_\mu$, and thus has
no consequence.  Also, we have equated $x = t,$ that is, set $c$ = 1.

\subsection{The Mikheyev-Smirnov-Wolfenstein Mechanism}

The view of neutrino oscillations changed when Mikheyev and
Smirnov~\cite{ms} showed in 1985 that the density dependence of the
neutrino effective mass, a phenomenon first discussed by Wolfenstein
in 1978, could greatly enhance oscillation probabilities: a $\nu_e$ is
adiabatically transformed into a $\nu_\mu$ as it traverses a critical
density within the sun.  It became clear that the sun was not only an
excellent neutrino source, but also a natural regenerator for cleverly
enhancing the effects of flavor mixing.
   
While the original work of Mikheyev and Smirnov was numerical, their
phenomenon was soon understood analytically as a level-crossing
problem.  If one writes the neutrino wave function in matter as in eq.
(\ref{eq:31}), the evolution of $a_e(x)$ and $a_\mu(x)$ is governed by
\begin{equation}
i {d \over dx} \left( \matrix { a_{\textstyle e} \cr
a_{\textstyle \mu} \cr} \right) = {1 \over 4E} \left ( \matrix{
2E \sqrt2 G_F \rho(x) - \delta m^2 \cos 2 \theta_{\textstyle v}
~~~~~\delta m^2\sin
2\theta_{\textstyle v} \cr 
\delta m^2\sin 2 \theta_{\textstyle v} ~~~ -2E \sqrt2 G_F \rho(x) +
\delta m^2
\cos 2\theta_{\textstyle v} \cr} \right) \left( \matrix {
a_{\textstyle e} \cr
a_{\textstyle \mu} \cr} \right) 
\end{equation}
where G$_F$ is the weak coupling constant and $\rho (x)$ the solar
electron density.  If $\rho (x)$ = 0, this is exactly our previous
result and can be trivially integrated to give the vacuum oscillation
solutions given above.  The new contribution to the diagonal elements,
$2 E \sqrt2 G_F \rho(x)$, represents the effective contribution to
the $M^2_\nu$ matrix that arises from neutrino-electron scattering.  The indices
of refraction of electron and muon neutrinos differ because the former
scatter by charged and neutral currents, while the latter have only
neutral current interactions.  The difference in the forward
scattering amplitudes determines the density-dependent splitting of
the diagonal elements of the matter equation,
the generalization of eq. (32).

It is helpful to rewrite this equation in a basis consisting of the
light and heavy local mass eigenstates (i.e., the states that
diagonalize the right-hand side of eq. (33)),
\begin{eqnarray}
|\nu_L (x)\rangle &=& \cos \theta (x)|\nu_e\rangle - \sin \theta 
(x)|\nu_\mu\rangle \nonumber \\
|\nu_H(x)\rangle &=& \sin \theta (x)|\nu_e\rangle + \cos \theta 
(x)|\nu_\mu \rangle . 
\end{eqnarray}
The local mixing angle is defined by
\begin{eqnarray}
\sin 2 \theta (x)  &=& {\sin 2 \theta_{\textstyle v} \over 
\sqrt{X^2 (x) + \sin^2
2\theta_{\textstyle v}}} \nonumber \\
\cos 2\theta (x)  &=& {-X (x) \over \sqrt{X^2 (x) + 
\sin^2 2\theta_{\textstyle v}}} 
\end{eqnarray}
where $X(x) = 2 \sqrt2G_F \rho(x) E/\delta m^2 - \cos
2\theta_{\textstyle v}$.  Thus $\theta(x)$ ranges from
$\theta_{\textstyle v}$ to $\pi/2$ as the density $\rho(x)$ goes from
0 to $\infty$.

If we define
\begin{equation}
|\nu (x) \rangle = a_H(x)|\nu_H(x)\rangle + a_L(x)|\nu_L(x)\rangle,
\end{equation}
the neutrino propagation can be rewritten in terms of the local
mass eigenstates
\begin{equation}
i {d \over dx} \pmatrix{
a_H \cr
a_L \cr} = \pmatrix {
\lambda(x) & i \alpha (x) \cr
-i \alpha (x) & - \lambda (x) \cr }
\pmatrix
{a_H \cr
a_L }
\end{equation}
with the splitting of the local mass eigenstates determined by
\begin{equation}
2 \lambda (x) = {\delta m^2 \over 2E} \sqrt{X^2 (x) + \sin^2 2 
\theta_{\textstyle v}} 
\end{equation}
and with mixing of these eigenstates governed by the density gradient
\begin{equation}
\alpha (x) = \left({E \over \delta m^2}\right)
 \, {\sqrt2 \, G_F {d \over dx}
\rho(x)
\sin 2 \theta_{\textstyle v} \over X^2 (x) + \sin^2 2 
\theta_{\textstyle v}}.
\end{equation}
The results above are quite interesting: the local mass eigenstates
diagonalize the matrix if the density is constant.  In such a limit,
the problem is no more complicated than our original vacuum
oscillation case, although our mixing angle is changed because of the
matter effects.  But if the density is not constant, the mass
eigenstates evolve as the density changes.  This is the crux
of the MSW effect.  Note that the splitting achieves its minimum
value, ${\delta m^2 \over 2E} \sin 2 \theta_v$, at a critical density
$\rho_c = \rho (x_c)$
\begin{equation}
2 \sqrt2 E G_F \rho_c = \delta m^2 \cos 2 \theta_v 
\end{equation}
that defines the point where the diagonal elements of the original
flavor matrix cross.

Our local-mass-eigenstate form of the propagation equation can be
trivially integrated if the splitting of the diagonal elements is
large compared to the off-diagonal elements (see eq. (37)),
\begin{equation}
\gamma (x) = \left|{\lambda (x) \over \alpha (x)}\right| = {\sin^2
2\theta_{\textstyle v} \over \cos
2\theta_{\textstyle v}} \, {\delta m^2 \over 2 E} \, {1 \over 
|{1 \over \rho_c}
{d \rho (x) \over
dx}|} {[X (x)^2 + \sin^2 2\theta_v]^{3/2} \over \sin^3 2\theta_v} 
\gg 1, 
\end{equation}
a condition that becomes particularly stringent near the crossing
point,
\begin{equation}
\gamma_c = \gamma (x_c) = {\sin^2 2\theta_v \over \cos 2\theta_v} 
\, {\delta
m^2 \over 2 E} \, {1 \over \left|{1 \over \rho_c} {d \rho (x) 
\over dx}|_{x = x_c}\right|} \gg 1. 
\end{equation}
That is, adiabaticity depends on the density scale height at the 
crossing point.  The resulting adiabatic electron neutrino survival
probability~\cite{bethe}, valid when $\gamma_c \gg 1$, is
\begin{equation}
P^{\rm adiab}_{\nu_e} = {1 \over 2} + {1 \over 2} \cos 2 \theta_v \cos 2
\theta_i 
\end{equation}
where $\theta_i = \theta (x_i)$ is the local mixing angle at the
density where the neutrino was produced.

\begin{figure}[htb]
\psfig{bbllx=1.2cm,bblly=2.0cm,bburx=18cm,bbury=14.5cm,figure=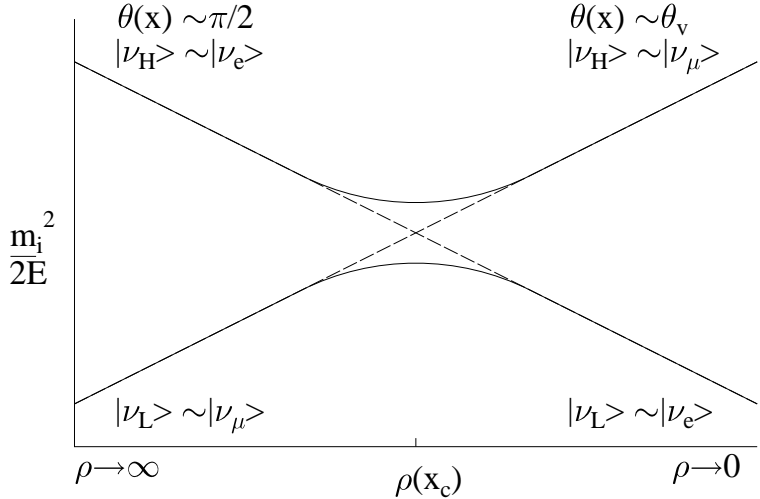,height=2.8in}
\caption{Schematic illustration of the MSW crossing.  The dashed 
lines correspond to the electron-electron and muon-muon diagonal
elements of the $M_\nu^2$ matrix in the flavor basis.  Their 
intersection defines the level-crossing density $\rho_c$.
The solid lines are the trajectories of the light and heavy
local mass eigenstates.  If the electron neutrino is produced 
at high density and propagates adiabatically, it will follow
the heavy-mass trajectory, emerging from the sun as a $\nu_\mu$.}
\end{figure}
  
The physical picture behind this derivation is illustrated in Figure
5.  One makes the usual assumption that, in vacuum, the $\nu_e$ is
almost identical to the light mass eigenstate, $\nu_L(0)$, i.e., $m_1
< m_2$ and $\cos \theta_v \sim$ 1.  But as the density increases, the
matter effects make the $\nu_e$ heavier than the $\nu_\mu$, with
$\nu_e \to \nu_H (x)$ as $\rho(x)$ becomes large.  The special
property of the Sun is that it produces $\nu_e$s at high density that
then propagate to the vacuum where they are measured.  The adiabatic
approximation tells us that if initially $\nu_e \sim \nu_H (x)$, the
neutrino will remain on the heavy mass trajectory provided the density
changes slowly.  That is, if the solar density gradient is
sufficiently gentle, the neutrino will emerge from the sun as the
heavy vacuum eigenstate, $ \sim \nu_\mu$.  This guarantees nearly
complete conversion of $\nu_e$s into $\nu_\mu$s, producing a flux that
cannot be detected by the Homestake or SAGE/GALLEX detectors.
   
But this does not explain the curious pattern of partial flux
suppressions coming from the various solar neutrino experiments.  The
key to this is the behavior when $\gamma_c \lsim$ 1.  Our expression
for $\gamma(x)$ shows that the critical region for non-adiabatic
behavior occurs in a narrow region (for small $\theta_v$) surrounding
the crossing point, and that this behavior is controlled by the
density scale height.  This suggests an analytic strategy for
handling non-adiabatic crossings: one can replace the true solar
density by a simpler (integrable!) two-parameter form that is
constrained to reproduce the true density and its derivative at the
crossing point $x_c$. Two convenient choices are the linear $(\rho (x)
= a + bx)$ and exponential $(\rho (x) = ae^{-bx})$ profiles.  As the
density derivative at $x_c$ governs the non-adiabatic behavior, this
procedure should provide an accurate description of the hopping
probability between the local mass eigenstates when the neutrino
traverses the crossing point.  The initial and ending points $x_i$ and
$x_f$ for the artificial profile are then chosen so that $\rho(x_i)$
is the density where the neutrino was produced in the solar core and
$\rho(x_f) = 0$ (the solar surface), as illustrated in in Figure 6.
Since the adiabatic result ($P_{\nu_e}^{\mathrm{adiab}}$) depends only
on the local mixing angles at these points, this choice builds in that
limit.  But our original flavor-basis equation can then be integrated
exactly for linear and exponential profiles, with the results given in
terms of parabolic cylinder and Whittaker functions, respectively.

\begin{figure}[htb]
\psfig{bbllx=0.0cm,bblly=2.8cm,bburx=16cm,bbury=21.3cm,figure=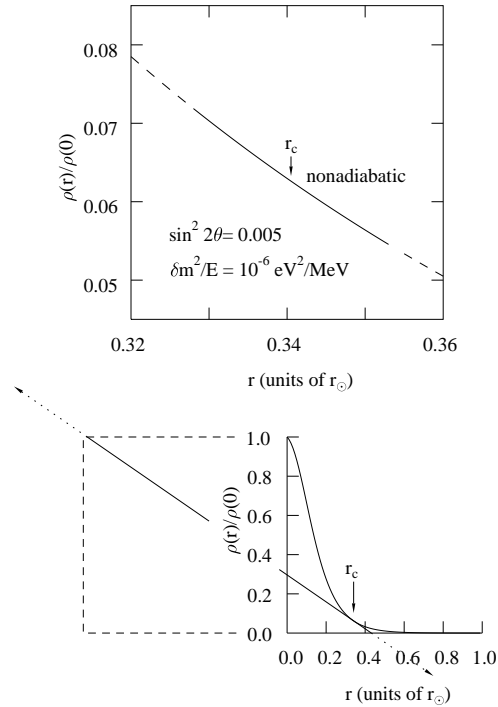,height=3.8in}
\caption{The top figure illustrates, for one choice of sin$^2 2\theta$
  and $\delta m^2$, that the region of non-adiabatic propagation (solid
  line) is usually confined to a narrow region around the crossing
  point $r_c$.  In the lower figure, the solid lines represent the
  solar density and a linear approximation to that density that has
  the correct initial and final values, as well as the correct density
  and density derivative at $r_c$.  Thus the linear profile is a very
  good approximation to the sun in the vicinity of the crossing point.
  The MSW equations can be solved analytically for this wedge.  By
  extending the wedge to $\pm \infty$ (dotted lines) and assuming
  adiabatic propagation in these regions of unphysical density, one
  obtains the simple Landau-Zener result discussed in the text.}
\end{figure}
  
That result can be simplified further by observing that the
non-adiabatic region is generally confined to a narrow region around
$x_c$, away from the endpoints $x_i$ and $x_f$.  We can then extend
the artificial profile to $x = \pm \infty$, as illustrated by the
dashed lines in Figure 6.  As the neutrino propagates adiabatically in
the unphysical region $x < x_i$, the exact solution in the physical
region can be recovered by choosing the initial boundary conditions
\begin{eqnarray}
a_L (- \infty) &=& - a_\mu (- \infty) = \cos \theta_i e^{- i 
\int^{x_i}_{- 
\infty} \lambda (x) dx} \nonumber\\
a_H (- \infty) &=& a_e (- \infty) = \sin \theta_i 
e^{i \int^{x_i}_{- \infty} 
\lambda (x) dx} . 
\end{eqnarray}
That is, $|\nu (-\infty)\rangle$ will then adiabatically evolve to
$|\nu (x_i)\rangle = |\nu_e\rangle$ as $x$ goes from $- \infty$ to
$x_i$.  The unphysical region $x > x_f$ can be handled similarly.

With some algebra a simple generalization of the adiabatic
result emerges that is valid for all $\delta m^2/E$ and $\theta_v$
\begin{equation}
P_{\nu_e} = {1 \over 2} + {1 \over 2} \cos 2 \theta_v \cos 2 
\theta_i ( 1 - 2P_{\rm {hop}}) 
\end{equation}
where P$_{\rm {hop}}$ is the Landau-Zener probability of hopping from
the heavy mass trajectory to the light trajectory on traversing the
crossing point.  For the linear approximation to the
density~\cite{hlz,plz},
\begin{equation}
P^{\rm {lin}}_{\rm {hop}} = e^{- \pi \gamma_c/2} . 
\end{equation}
As it must by our construction, $P_{\nu_e}$ reduces to P$^{\rm
  {adiab}}_{\nu_e}$ for $\gamma_c \gg$ 1.  When the crossing becomes
non-adiabatic (e.g., $\gamma_c \ll 1$ ), the hopping probability goes
to 1, allowing the neutrino to exit the sun on the light mass
trajectory as a $\nu_e$, i.e., no conversion occurs.

Thus there are two conditions for strong conversion of solar
neutrinos: there must be a level crossing (that is, the solar core
density must be sufficient to render $\nu_e \sim \nu_H (x_i)$ when it
is first produced) and the crossing must be adiabatic.  The first
condition requires that $\delta m^2/E$ not be too large, and the
second $\gamma_c \gsim$ 1.  The combination of these two constraints,
illustrated in fig. 7, defines a triangle of interesting parameters in
the ${\delta m^2 \over E} - \sin^2 2\theta_v$ plane, as Mikheyev and
Smirnov found by numerically integration.  A remarkable feature of
this triangle is that strong $\nu_e \to \nu_\mu$ conversion can occur
for very small mixing angles $(\sin^2 2 \theta \sim10^{-3}$), unlike
the vacuum case.

\begin{figure}[htb]
\psfig{bbllx=-1.5cm,bblly=0.0cm,bburx=15cm,bbury=22.0cm,figure=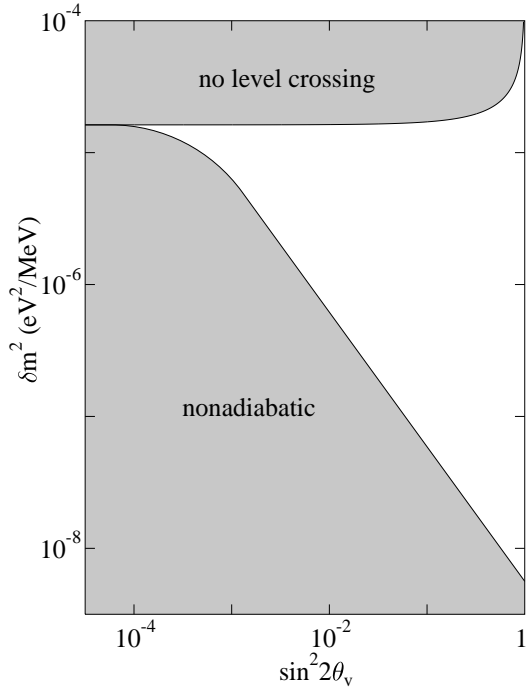,height=3.8in}
\caption{MSW conversion for a neutrino produced at the sun's
center.  The upper shaded region indicates those $\delta m^2/E$
where the vacuum mass splitting is too great to be overcome
by the solar density.  Thus no level crossing occurs.  The
lower shaded region defines the region where the level crossing
is non-adiabatic ($\gamma_c$ less than unity).  The unshaded
region corresponds to adiabatic level crossings where strong
$\nu_e \rightarrow \nu_\mu$ will occur.}
\end{figure}
  
One can envision superimposing on fig. 7 the spectrum of solar
neutrinos, plotted as a function of ${\delta m^2 \over E}$ for some
choice of $\delta m^2$.  Since Davis sees {\it some} solar neutrinos,
the solutions must correspond to the boundaries of the triangle in
fig. 7.  The horizontal boundary indicates the maximum ${\delta m^2
  \over E}$ for which the sun's central density is sufficient to cause
a level crossing.  If a spectrum properly straddles this boundary, we
obtain a result consistent with the Homestake experiment in which low
energy neutrinos (large 1/E) lie above the level-crossing boundary
(and thus remain $\nu_e$'s), but the high-energy neutrinos (small 1/E)
fall within the unshaded region where strong conversion takes place.
Thus such a solution would mimic nonstandard solar models in that only
the $^8$B neutrino flux would be strongly suppressed.  The diagonal
boundary separates the adiabatic and non-adiabatic regions.  If the
spectrum straddles this boundary, we obtain a second solution in which
low energy neutrinos lie within the conversion region, but the
high-energy neutrinos (small 1/E) lie below the conversion region and
are characterized by $\gamma \ll 1$ at the crossing density.  (Of
course, the boundary is not a sharp one, but is characterized by the
Landau-Zener exponential).  Such a non-adiabatic solution is quite
distinctive since the flux of pp neutrinos, which is strongly
constrained in the standard solar model and in any steady-state
nonstandard model by the solar luminosity, would now be sharply
reduced.  Finally, one can imagine ``hybrid" solutions where the
spectrum straddles both the level-crossing (horizontal) boundary and
the adiabaticity (diagonal) boundary for small $\theta$, thereby
reducing the $^7$Be neutrino flux more than either the pp or $^8$B
fluxes.

\begin{figure}[htb]
  \vspace{8pt} \centerline{\hbox{\epsfxsize=3.0 in 
\epsfbox[125 139 485 653]{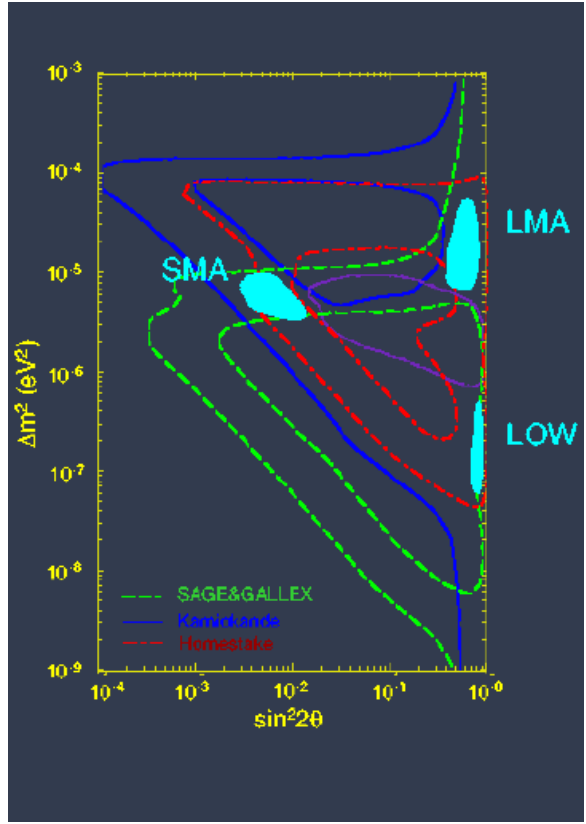}}}
\caption{Illustration of the SMA, LMA, and LOW solutions arising out
of fits to the event rates for Cl, Kamiokande, and SAGE/GALLEX.  Figure
provided by K. Heeger.}
\end{figure}
  
What are the results of a careful search for MSW solutions satisfying
the Homestake, Kamiokande, and SAGE/GALLEX
constraints?  This was explored in detail by several groups (see fig. 8).  One
solution, corresponding to a region surrounding $\delta m^2 \sim 6
\cdot 10^{-6} $eV$^2$ and $\sin^2 2\theta_v \sim 6 \cdot 10^{-3}$, is
the hybrid case described above.  It is commonly called the
small-mixing-angle solution (SMA).  A second, large-angle solution (LMA) 
corresponds to $\delta m^2 \sim 10^{-5} $eV$^2$ and $\sin^2 2
\theta_v \sim$ 0.6.  These solutions can be distinguished by their
characteristic distortions of the solar neutrino spectrum.  The
survival probabilities $P_{\nu_e}^{\rm MSW}$(E) for the small- and
large-angle parameters given above are shown as a function of E in
fig. 9.

\begin{figure}[htb]
\psfig{bbllx=0.5cm,bblly=1.3cm,bburx=18cm,bbury=13.7cm,figure=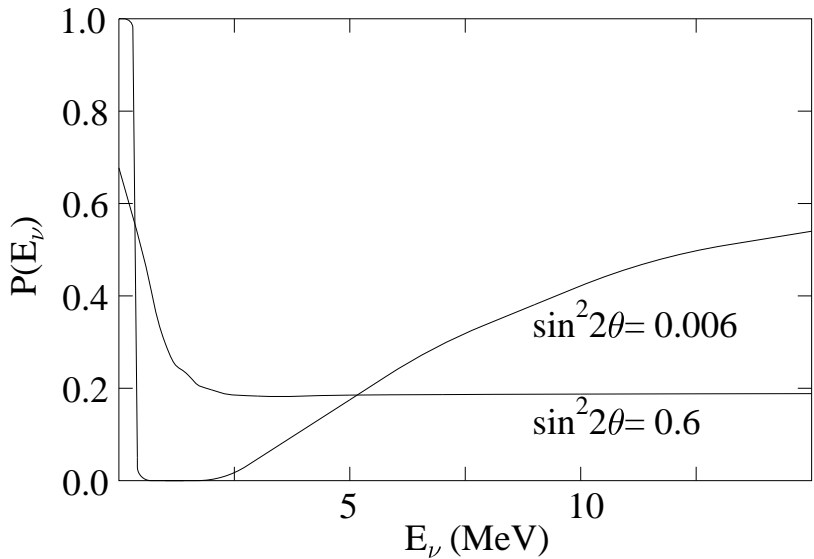,height=3.0in}
\caption{MSW survival probabilities P(E$_\nu$) for typical small
angle and large angle solutions.}
\end{figure}
  
The MSW mechanism provides a natural explanation for the pattern of
observed solar neutrino fluxes.  While it requires profound new
physics, both massive neutrinos and neutrino mixing are expected in
extended models.

\subsection{SuperKamiokande, SNO, and the Neutrino Mixing Matrix}

Over the past five years the Cl/SAGE/GALLEX/Kamiokande hints of 
new neutrino physics have been spectacularly confirmed by 
SuperKamiokande and SNO, and a program to further constrain the
neutrino mass matrix with terrestrial neutrino experiments has
been inaugurated by KamLAND and K2K.
  
SuperKamiokande and Sudbury Neutrino Observatory (SNO) detectors are
real-time counting detectors, in contrast to the radiochemical
detectors such as Homestake, GALLEX/GNO, and SAGE, which can only
determine a time- and energy-integral of the flux.
Both SuperKamiokande and SNO can detect neutrinos through
elastic scattering
\begin{equation}
  \label{eq:sno1}
   \nu_x + e^- \rightarrow \nu_x + e^-.
\end{equation}
The electrons coming from this reaction are confined to a forward
cone, with most of the observed spread in angles around the forward
peak coming from the limited angular resolution of the detector.
To an excellent approximation the recorded events (including the effects of resolution) are  
contained in the forward hemisphere.  Thus experimentalist can use
the isotropic distribution of events in the backward hemisphere
(defined around the vector pointing from the sun)
to determine the background, then substract in the forward 
hemisphere to obtain the events coming from solar neutrinos.
The neutrino energy is difficult to reconstruct 
because the initial neutrino momentum is shared by the scattered
neutrino and electron.  Nevertheless, for MSW solutions like the
SMA where there is a significant distortion of the $\nu_e$ 
spectrum, there is a sufficient residual distortion of the 
electron spectrum to signal new physics.  The SuperKamiokande
collaboration carefully calibrated the detector with electrons
from a linac so that small spectral distortions could be reliably
extracted.

In addition to the reaction eq. (\ref{eq:sno1}) SNO can detect
neutrinos by two additional reactions, one via the charged current
\begin{equation}
  \label{eq:sno2}
  \nu_e + \mathrm{d} \rightarrow \mathrm{p} + \mathrm{p} + e^-, 
\end{equation}
and the second via neutral current scattering
\begin{equation}
  \label{eq:sno3}
  \nu_x (\overline{\nu}_x)+ \mathrm{d} \rightarrow \nu_x (\overline{\nu}_x)+ \mathrm{p}
  + \mathrm{n}. 
\end{equation}
The neutrons produced in eq. (\ref{eq:sno3}) can be detected either
by (n,$\gamma)$ on the heavy water or on a salt introduced to
enhance the capture, or
by using $^3$He proportional counters. The electrons coming from the
reaction (\ref{eq:sno2}) are quite hard, with energies not too 
different from
$\sim E_{\nu} - 1.44$ MeV, and with a angular distribution 
approximately that of a pure GT transition in the relativistic
limit, ($1- \cos \theta_e /3$) with respect to the incident
neutrino.  This backward peaking contrasts nicely with the 
forward-peaked elastic scattering signal.

The hardness of the charged current reaction (eq. (48)) makes it
an effective tool for finding spectral distortions induced by the
MSW mechanism.  The SNO spectral distortion expect for the SMA
solution ($\delta m^2 \sim 5
\times 10^{-6}$ eV$^2$ and $\sin 2 \theta \sim 0.01$) is shown in
fig. 10.
  
\begin{figure}[t]
  \vspace{8pt} \centerline{\hbox{\epsfxsize=3 in \epsfbox[19 65 526
      685]{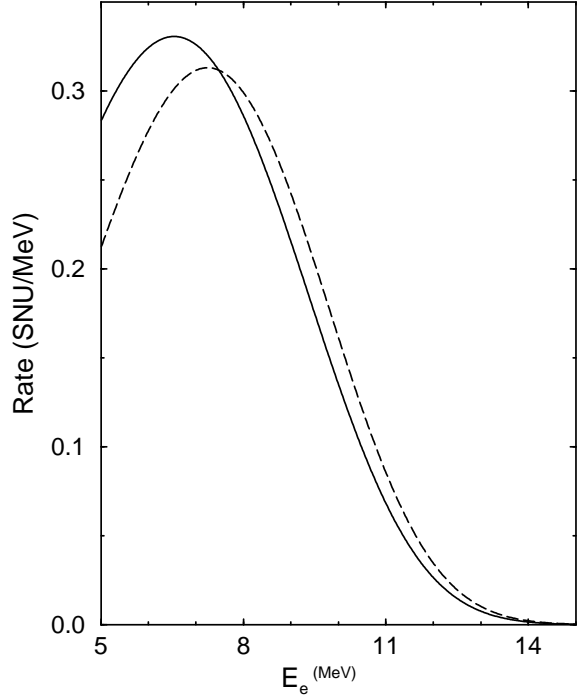}}}
\caption{The dashed line shows the spectrum distortion at SNO for 
  the small-angle MSW solution ($\delta m^2 \sim 5 \times 10^{-6}$
  eV$^2$ and $\sin 2 \theta \sim 0.01$). The solid line is the
  spectrum without MSW oscillations, normalized to the same total rate
  as with MSW oscillations.} 
\vspace{8pt}
\end{figure}

If there are no oscillations into sterile states (new neutrino 
states lacking the usual standard model weak interactions),
the neutral current reaction (eq. (49)) measures the total SSM
flux, independent of flavor.  One other reaction of potential
interest
\begin{equation}
 \bar{\nu}_e + \mathrm{d} \rightarrow \mathrm{n} + \mathrm{n} + e^+
\end{equation}
produces a two-neutron coincidence.  Electron antineutrinos can arise
from spin-flavor oscillations~\cite{lim,ak} in the sun and, of course, from supernovae.

As the solar neutrino problem deepened with the new measurements
from SAGE/GALLEX and Kamiokande, a second neutrino anomaly also
began to draw attention.  Several early underground detectors
(IMB, Kamiokande, and later Soudan II) found a deficit of muons
produced by atmospheric neutrinos.
Atmospheric neutrinos arise from the decay of secondary pions, kaons,
and muons produced by the collisions of primary cosmic rays with the
oxygen and nitrogen nuclei in the upper atmosphere. For energies less
than 1 GeV all the secondaries decay :
\begin{eqnarray}
\pi^{\pm} (K^{\pm}) &\rightarrow &\mu^{\pm} + \nu_{\mu}
(\overline{\nu}_{\mu}), \nonumber\\ \mu^{\pm} &\rightarrow & e^{\pm} +
\nu_e (\overline{\nu}_e) +  \overline{\nu}_{\mu} (\nu_{\mu}).
\end{eqnarray}
Consequently one expects the ratio
\begin{equation}
r = (\nu_e + \overline{\nu}_e) / (\nu_{\mu} + \overline{\nu}_{\mu})
\end{equation}
to be approximately 0.5 in this energy range. Detailed Monte Carlo
calculations \cite{gaisser}, including the effects of muon
polarization, give $  r \sim 0.45$. As one is evaluating a ratio of
similarly calculated processes, $r$ does not require an absolute
flux calculation and is thus relative free of systematic 
uncertainties. Different groups estimating this ratio, even though they
start with neutrino fluxes which can differ in magnitude by up to
25\%, all agree within a few percent \cite{bludman}. As the shower
energy increases more muons survive due to time dilation. Hence one
expects the ratio $r$ to decrease as the energy increases. The ratio
(observed to predicted) of ratios
\begin{equation}
R = {(\nu_{\mu} / \nu_e)_{\rm data} \over (\nu_{\mu} / \nu_e)_{\rm
Monte  Carlo} }
\end{equation}
studied by the experimentalists
should then be unity, in the absence of oscillations.

The first ``smoking gun'' for neutrino oscillations came from 
the detailed atmospheric $\nu$ results of SuperKamiokande.
The initial announcement of neutrino oscillations, made in 1998,
is now supported by 1489 days of data.  The experimenters found
\begin{equation}
  \label{atm1}
  R=0.638 \pm 0.016 ({\rm stat}) \pm 0.050 ({\rm syst}) \nonumber
\end{equation}
for sub-GeV events which were fully contained in the detector and
\begin{equation}
  \label{atm2}
  R=0.658^{+0.030}_{-0.028} ({\rm stat}) \pm 0.078 ({\rm syst}) \nonumber
\end{equation}
for fully- and partially-contained multi-GeV events. The large
deviation from $R=1$ indicates neutrino
oscillations.  Much more dramatic evidence
for oscillations comes from measuring $R$ as a function of the
zenith angle, $\Theta$, between the vertical and incident neutrino. A
down-going neutrino ($\Theta \sim 0^o$) travels through the atmosphere
above the detector (a distance of about 20 km), whereas an up-going
neutrino ($\Theta \sim 180^o$) has traveled through the entire Earth
(a distance of about 13000 km). Hence a measurement of the flux
as a function of the zenith angle yields information about
neutrino survival probabilities as a function of the distance traveled. 

\begin{figure}[t]
  \vspace{8pt} \centerline{\hbox{\epsfxsize=3.6 in 
\epsfbox[8 -3 502 485]{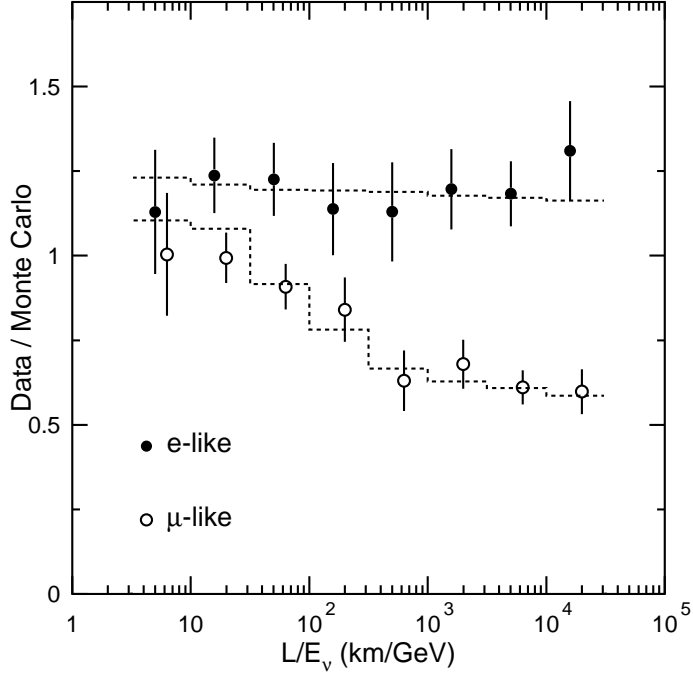}}}
\caption{The ratio of fully contained events measured
  at SuperKamiokande versus reconstructed $L/E_{\nu}$.  The dashed
  lines show the expected shape for $\nu_{\mu} \leftrightarrow
  \nu_{\tau}$ oscillations with $\delta m^2 = 2.2 \cdot 10^{-3}$
  eV$^2$ and $\sin^2 2 \theta = 1$. } 
\vspace{8pt}
\end{figure}

The SuperKamiokande collaboration measured the zenith angle dependence
not only of $R$, but also of the electron and muon neutrino fluxes
separately \cite{skatm}. This information is shown in Fig. 11, where
the data is plotted as a function of the reconstructed $L/E_{\nu}$
instead of the zenith angle. The data exhibit a zenith-angle
(distance) dependent deficit of muon neutrinos, but not of electron
neutrinos, a behavior consistent with $\nu_{\mu} \rightarrow
  \nu_{\tau}$ oscillations. This interpretation is consistent with the
  deficits of up-going muons measured with the Kamiokande \cite{katm} and
  MACRO \cite{matm} detectors. These muons are produced by the very
  high energy up-going
muon neutrinos in the rock surrounding the detectors.

One remarkable feature of the SuperKamiokande atmospheric neutrino
results is the deduced mixing angle.  Designating the two mass 
eigenstates participating in the mixing as 2 and 3,
the best fit $\sin^2 2\theta_{23}$
is 1, with $\sin^2\theta_{23} \ge 0.92$ at 90\% c.l.  As quark
mixing angles are small, a $\nu$ mixing angle near maximal
($\theta_{23} \sim 45^\circ$) was not
anticipated.  The best-fit $\delta m^2_{23}$ is $\sim 2.5 \cdot 10^{-3}$
eV$^2$.

This year an equally spectacularly resolution of the solar neutrino
problem was obtained by SNO.  SNO's proof of oscillations was
definitive: the heavy-flavor neutrinos produced as
a result of solar $\nu_e$ oscillations were seen directly,
by comparing the charge current and neutral current results.

SNO's initial results are shown in fig. 12.  The detector's three
$\nu$ channels -- charge and neutral current scattering off 
deuterium and $\nu$-electron elastic scattering -- combine to
show that approximately two-thirds of the solar flux is in
heavy flavors.  The fluxes deduced, assuming a standard $^8$B $\nu$ spectrum shape, are~\cite{SNO}
\begin{eqnarray}
\phi_{\nu_e} = (1.76 \pm 0.05 (stat) \pm 0.09 (syst)) \cdot 10^6/\mathrm{cm}^2 \mathrm{s} \nonumber \\
\phi_{\nu_{heavy}} = (3.41 \pm 0.45 (stat) {}^{+0.48}_{-0.45} (syst)) \cdot 10^6/\mathrm{cm}^2 \mathrm{s}. 
\end{eqnarray}
The 5.3 $\sigma$ difference is the significance of the oscillation
proof.  The total active flux, measured above the neutron breakup
threshold for deuterium of 2.2 MeV, is 
\begin{equation}
\phi_{\nu} = 5.09^{+0.44}_{-0.43} (stat) {}^{+0.46}_{-0.43}) \cdot 10^6/\mathrm{cm}^2 \mathrm{s}
\end{equation}
in excellent agreement with SSM results.  

Furthermore the results point to a unique solution in the MSW
plane, the LMA solution, with a best fit $\delta m^2 \sim 5 \cdot
10^{-5}$ eV$^2$ and $\sin^2 2\theta \sim 0.75$.  Thus the oscillation
corresponds to a new $\delta m^2$, distinct from the atmospheric
$\nu$ solution, and while the oscillation appears to be not quite
maximal, the mixing angle is again large.
For three light neutrinos, we can label this mixing as that 
between mass eigenstates 1 and 2 with $\theta_{12} \sim 30^\circ$.

\begin{figure}[t]
  \vspace{8pt} \centerline{\hbox{\epsfxsize=4.5 in 
\epsfbox[0 0 565 405]{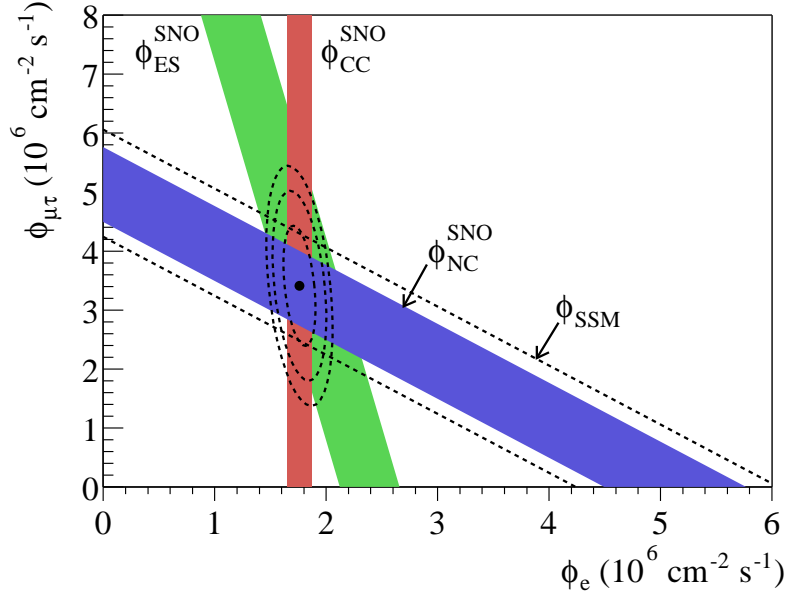}}}
\caption{The $^8$B solar neutrino flux decomposed into electron
and heavy-flavor components.  The diagonal bands show the total
flux measured by the SNO neutral-current reaction (solid) and
predicted by the SSM (dashed).  The intersection with charge-current
and neutral current bands determines the flavor content. From
Ref. [8].}
\vspace{8pt}
\end{figure}

\begin{figure}[t]
  \vspace{8pt} \centerline{\hbox{\epsfxsize=3.6in 
\epsfbox[0 0 566 520]{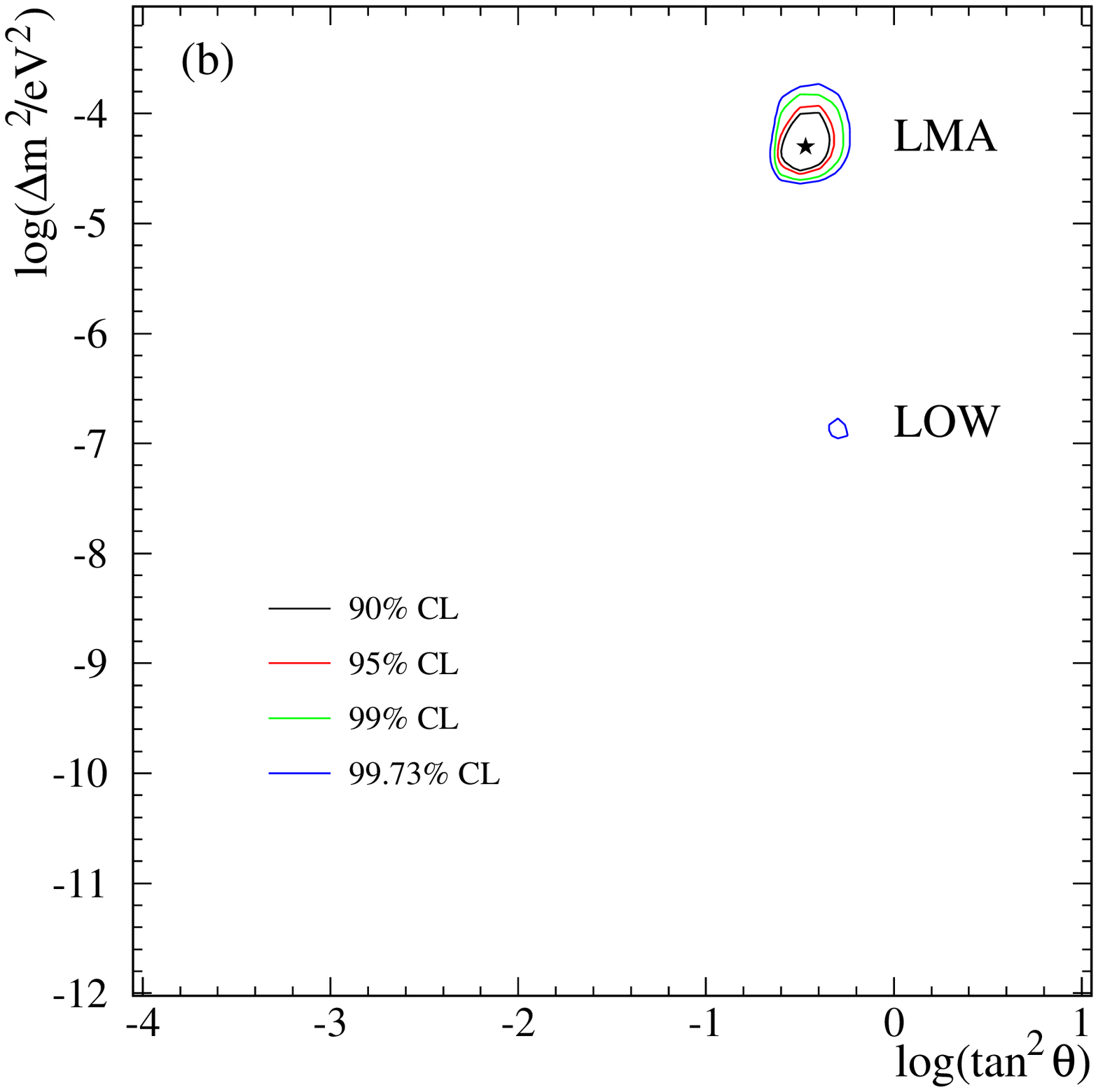}}}
\caption{Allowed regions of the MSW plane consistent with the SNO
day and night energy spectra, the results from the Cl and 
SAGE/GALLEX/GNO experiments, and the day and night spectra from
SuperKamiokande.  The star indicates the best fit.  Only the LMA
region is allowed at 99\% c.l.  From Ref. [32].}
\vspace{8pt}
\end{figure}

The SuperKamiokande and SNO determinations of $\theta_{23}$ and
$\theta_{12}$ are an exciting and important step in defining
the neutrino mixing matrix, a quantity we hope will point the
way to the next standard model.
The three-flavor neutrino mixing matrix is conventionally written
as
\begin{eqnarray}
 \left( \begin{array}{c} \nu_e \\ \nu_\mu \\ \nu_\tau \end{array} \right)
&=& \left( \begin{array}{ccc} c_{12}c_{13} & s_{12}c_{13} & s_{13} e^{-i \delta} \\
-s_{12}c_{23}-c_{12}s_{23}s_{13} e^{i \delta} & c_{12}c_{23}-s_{12}s_{23}s_{13} e^{i \delta} & s_{23}c_{13} \\
s_{12}s_{23}-c_{12}c_{23}s_{13} e^{i \delta} & -c_{12}s_{23}-s_{12}c_{23}s_{13} e^{i \delta} & c_{23}c_{13} 
\end{array} \right) 
\left( \begin{array}{c} \nu_1 \\ \nu_2 \\ \nu_3 \end{array} \right) \nonumber \\
 &=& \left( \begin{array}{ccc} 1 & & \\ & c_{23} & s_{23} \\ & -s_{23} & c_{23} \end{array} \right)
\left( \begin{array}{ccc} c_{13} & & s_{13} e^{-i \delta} \\ & 1 & \\ -s_{13} e^{i \delta} & & c_{13} \end{array} \right)
\left( \begin{array}{ccc} c_{12} & s_{12} & \\ -s_{12} & c_{12} & \\ & & 1 \end{array} \right)
\left( \begin{array}{c} \nu_1 \\ \nu_2 \\ \nu_3 \end{array} \right) 
\end{eqnarray}
Here $c_{12} = \cos \theta_{12}$, etc.  Despite the discoveries
of the past five years, we have yet to complete this matrix.
Mixing between eigenstates 1 and 3 has not yet measured: disappearance results
from the Chooz~\cite{chooz} and Palo Verde~\cite{pv} reactor experiments limit $\theta_{13} \lsim 10^\circ$
in the atmospheric $\delta m^2$ range.
Furthermore there is great interest in measuring CP violation effects,
due to the large mixing angles so far measured and to 
the possibility that the baryon number asymmetry arises through
leptogenesis.  CP violation in oscillations requires a nonvanishing
$\sin \theta_{13}$ in addition to $\delta$.  Several very long baseline oscillation 
proposals have been made in which CP-violation
would be disentangled from matter effects and 
from various neutrino parameter uncertainties.

Recently two important laboratory neutrino oscillation results 
have added to our picture of the mixing matrix.  
The atmospheric neutrino $\delta m^2$ range has been tested in the
K2K experiment, in which events initiated by $\sim$ 1 GeV $\nu_\mu$s produced by the KEK 
proton synchrontron are recorded
in SuperKamiokande~\cite{K2K}.  The null hypothesis of no oscillations is 
allowed only at a confidence level $\lsim 0.007$.  The best-fit
oscillation parameters, $\delta m^2 \sim 2.7 \cdot 10^{-3}$ eV$^2$
and $\sin^2 2 \theta \sim 1$, are in excellent agreement with the
atmospheric neutrino values.  The current data represents about 
half of the beam time expected for the experiment.

Very recently KamLAND, the first terrestrial experiment to probe
the solar $\delta m^2$ range, announced initial results~\cite{KamLAND}.  
This long-baseline experiment records reactor $\bar{\nu}_e$
events in a liquid scintillator detector built at the site
that once housed Kamiokande.  Initial results
confirm oscillations at 99.95\% c.l., rule out all two-flavor
solar neutrino solutions other than LMA, and significantly
narrow the range of allowed $\delta m_{12}^2$
when combined with the solar neutrino results.  The 
KamLAND results are shown in fig. 14.
Ultimately results from KamLAND combined with a high-precision
measurement of solar pp neutrinos could determine both 
$\delta m_{12}^2$ and $\sin^2 2\theta_{12}$ 
with increased accuracy.

\begin{figure}[t]
  \vspace{8pt} \centerline{\hbox{\epsfxsize=3.6in 
\epsfbox[0 0 570 539]{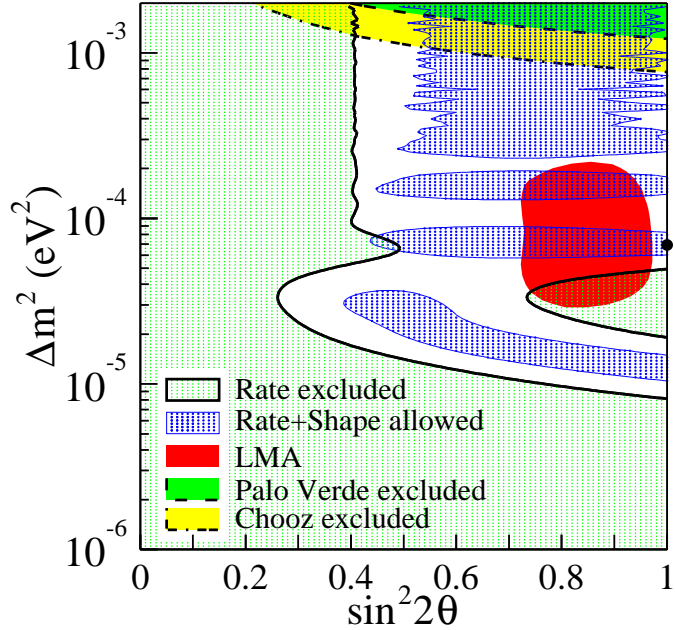}}}
\caption{Two-neutrino oscillation constraints from KamLAND 
superimposed on the solar neutrino LMA allowed region.
From Ref. [36].}
\vspace{8pt}
\end{figure}

Important question remain about the masses as well.  All of the
oscillation results test mass differences.  Thus the absolute
scale of $\nu$ mass must be probed in other experiments, such as
tritium $\beta$ decay.  The current limit from the Mainz and Troitsk  
experiments is $m_{\bar{\nu}_e} \lsim 2.2$ eV \cite{wein}.  As results from large-scale structure surveys
now underway will be affected by $\nu$ masses
as small as 0.3 eV, further progress must be made in direct
$\nu$ mass experiments to eliminate a potentially
significant cosmological uncertainty.

We do not know the mass hierarchy.  Both normal (the solar
neutrino mass pair light) and inverted (this pair heavy) are
allowed by the data.

We do not know the charge conjugation properties of neutrinos.
Neutrinos are unique among standard model fermions in allowing
both Dirac and Majorana mass terms, as was noted earlier in the
discussion of the seesaw mechanism.  These mass terms
can be distinguished because
Majorana masses break lepton number conservation, allowing
neutrinoless $\beta \beta$ decay to take place.  Several recent
next-generation $\beta \beta$ decay proposals argue that
current lifetime limits can be improved by factors of $\sim$ 100,
yielding sensitivities to Majorana masses of 10-50 milli-eV.
Many mass scenarios consistent with the solar and atmospheric
neutrino results predict neutrinoless $\beta \beta$ decay at
this level.

Thus we are a very exciting juncture.  We have found the first 
evidence for physics beyond the standard model, and we know this
physics has immediate consequences beyond nuclear and particle
physics, affecting particle dark matter, large-scale
structure, and baryogenesis.  The discoveries coming from 
astrophysical neutrinos are stimulating ambitious new experiments
with both terrestrial and astrophysical beams.  A great deal of
information remains hidden and yet is accessible to the next
generation of experiments, experiments that will  
require new neutrino beams and long-baseline megadetectors,
as well as high sensitive instruments for measuring double
beta decay and probing the low-energy portion of the solar 
neutrino flux.  There are many reasons to hope that these 
endeavors will provide an experimental foundation for constructing
the next standard model of subatomic physics.
  
\section{Supernovae, Supernova Neutrinos, and Nucleosynthesis}

Consider a massive star, in excess of 10 solar masses, burning the
hydrogen in its core under the conditions of hydrostatic equilibrium.
When the hydrogen is exhausted, the core contracts until the density
and temperature are reached where 3$\alpha \rightarrow ^{12}$C can
take place.  The He is then burned to exhaustion.  This pattern (fuel
exhaustion, contraction, and ignition of the ashes of the previous
burning cycle) repeats several times, leading finally to the explosive
burning of $^{28}$Si to Fe.  For a heavy star, the evolution is rapid:
the star has to work harder to maintain itself against its own
gravity, and therefore consumes its fuel faster.  A 25 solar mass star
would go through all of these cycles in about 7 My, with the final
explosive Si burning stage taking a few days.  The result is an
``onion skin" structure of the pre-collapse star in which the star's
history can be read by looking at the surface inward: there are
concentric shells of H, $^4$He, $^{12}$C, $^{16}$O and $^{20}$Ne,
and $^{28}$Si, with Fe at the center.

\subsection{The Explosion Mechanism~\protect\cite{mezz}}
The source of energy for this evolution is nuclear binding energy.  A
plot of the nuclear binding energy $\delta$ as a function of nuclear
mass shows that the minimum is achieved at Fe.  In a scale where the
$^{12}$C mass is picked as zero:
\begin{center}
$^{12}$C~~~~~$\delta$/nucleon = 0.000 MeV \\
$^{16}$O~~~~~$\delta$/nucleon = -0.296 MeV \\
$^{28}$Si~~~~$\delta$/nucleon = -0.768 MeV \\
$^{40}$Ca~~~~$\delta$/nucleon = -0.871 MeV \\
$^{56}$Fe~~~~$\delta$/nucleon = -1.082 MeV \\
$^{72}$Ge~~~~$\delta$/nucleon = -1.008 MeV \\
$^{98}$Mo~~~~$\delta$/nucleon = -0.899 Mev
\end{center}
Once the Si burns to produce Fe, there is no further source of
nuclear energy adequate to support the star.  So as the last remnants
of nuclear burning take place, the core is largely supported by
degeneracy pressure, with the energy generation rate in the core being
less than the stellar luminosity.  The core density is about 2 $\times
10^9$ g/cc and the temperature is kT $\sim$ 0.5 MeV.

Thus the collapse that begins with the end of Si burning is not halted
by a new burning stage, but continues.  As gravity does work on the
matter, the collapse leads to a rapid heating and compression of the
matter.  As the nucleons in Fe are bound by about 8 MeV, sufficient
heating can release $\alpha$s and a few nucleons.  At the same time,
the electron chemical potential is increasing.  This makes electron
capture on nuclei and any free protons favorable,
\begin{equation}
 e^- + p \rightarrow \nu_e + n. 
\end{equation}
Note that the chemical equilibrium condition is
\begin{equation}
 \mu_e + \mu_p = \mu_n + \langle E_\nu \rangle. 
\end{equation}
Thus the fact that neutrinos are not trapped plus the rise in the
electron Fermi surface as the density increases, lead to increased
neutronization of the matter.  The escaping neutrinos carry off energy
and lepton number.  Both the electron capture and the nuclear
excitation and disassociation take energy out of the electron gas,
which is the star's only source of support.  This means that the
collapse is very rapid.  Numerical simulations find that the iron core
of the star ($\sim$ 1.2-1.5 solar masses) collapses at about 0.6 of the
free fall velocity.

In the early stages of the infall the $\nu_e$s readily escape.  But
neutrinos are trapped when a density of $\sim$ 10$^{12}$g/cm$^3$ is
reached.  At this point the neutrinos begin to scatter off the matter
through both charged current and coherent neutral current processes.
The neutral current neutrino scattering off nuclei is particularly
important, as the scattering cross section is off the total nuclear
weak charge, which is approximately the neutron number.  This process
transfers very little energy because the mass energy of the nucleus is
so much greater than the typical energy of the neutrinos.  But
momentum is exchanged.  Thus the neutrino ``random walks" out of the
star.  When the neutrino mean free path becomes sufficiently short,
the ``trapping time" of the neutrino begins to exceed the time scale
for the collapse to be completed.  This occurs at a density of about
10$^{12}$ g/cm$^3$, or somewhat less than 1\% of nuclear density.
After this point, the energy released by further gravitational
collapse and the star's remaining lepton number are trapped within the
star.

If we take a neutron star of 1.4 solar masses and a radius of
10 km, an estimate of its binding energy is
\begin{equation}
 {G M^2 \over 2R} \sim 2.5 \times 10^{53} \mathrm{ergs}. 
\end{equation}
Thus this is roughly the trapped energy that later will be radiated in
neutrinos.

The trapped lepton fraction $Y_L$ is a crucial parameter in the
explosion physics: a higher trapped $Y_L$ leads to a larger homologous
core, a stronger shock wave, and easier passage of the shock wave
through the outer core, as will be discussed below.  Most of the
lepton number loss of an infalling mass element occurs as it passes
through a narrow range of densities just before trapping.  The reasons
for this are relatively simple: on dimensional grounds weak rates in a
plasma go as $T^5$, where T is the temperature.  Thus the electron
capture rapidly turns on due to heating of the matter as it falls toward the trapping radius,
and lepton number loss is maximal just prior to trapping.  Inelastic
neutrino reactions have an important effect on these losses, as the
coherent trapping cross section goes as $E_\nu^2$ and is thus least
effective for the lowest energy neutrinos.  As these neutrinos escape,
inelastic reactions repopulate the low energy states, allowing the
neutrino emission to continue.

The velocity of sound in matter rises with increasing density.  The
inner homologous core, with a mass $M_{HC} \sim 0.6-0.9 $ solar
masses, is that part of the iron core where the sound velocity exceeds
the infall velocity.  In this subsonic central region any pressure variations that may
develop in the homologous core during infall can smooth out before the
collapse is completed.  As a result, the homologous core collapses as
a unit, retaining its density profile.  That is, if nothing were to
happen to prevent it, the homologous core would collapse to a point.

The collapse of the homologous core continues until nuclear densities
are reached.  As nuclear matter is rather incompressible ($\sim$ 200
MeV/f$^3$), the nuclear equation of state is effective in halting the
collapse: maximum densities of 3-4 times nuclear are reached, e.g.,
perhaps $6 \cdot 10^{14}$ g/cm$^3$.  The innermost shell of matter
reaches this supernuclear density first, rebounds, sending a pressure
wave out through the homologous core.  This wave travels faster than
the infalling matter, as the homologous core is characterized by a
sound speed in excess of the infall speed.  Subsequent shells follow.
The resulting series of pressure waves collect near the sonic point
(the edge of the homologous core).  As this point reaches nuclear
density and comes to rest, a shock wave breaks out and begins its
traversal of the outer core.

Initially the shock wave may carry an order of magnitude more energy
than is needed to eject the mantle of the star (less than 10$^{51}$
ergs).  But as the shock wave travels through the outer iron core, it
heats and melts the iron that crosses the shock front, at a loss of
$\sim$ 8 MeV/nucleon.  The enhanced electron capture that occurs off
the free protons left in the wake of the shock, coupled with the
sudden reduction of the neutrino opacity of the matter (recall
$\sigma_{coherent} \sim N^2$), greatly accelerates neutrino emission.
This is another energy loss.\footnote{Many numerical models predict,
  in conjunction with this sudden decrease in opacity, a
  strong ``breakout" burst of $\nu_e$s in the few milliseconds
  required for the shock wave to travel from the edge of the
  homologous core to the neutrinosphere at $\rho \sim 10^{12}$
  g/cm$^3$ and $r \sim 50$ km.  The neutrinosphere is the term for
  the neutrino trapping radius, or surface of last scattering.}  The
summed losses from shock wave heating and neutrino emission are
comparable to the initial energy carried by the shock wave.  Thus most
numerical models fail to produce a successful ``prompt" hydrodynamic
explosion: the shock stalls before it reaches the outer mantle.

Most of the attention in the past decade focused on two explosion
scenarios.  In the prompt mechanism described above, the shock wave is
sufficiently strong to survive the passage of the outer iron core with
enough energy to blow off the mantle of the star.  The most favorable
results were achieved with smaller stars (less than 15 solar masses)
where there is less overlying iron, and with soft equations of state,
which produce a more compact neutron star and thus lead to more energy
release.  In part because of the lepton number loss problems discussed
earlier, now it is widely believed that this mechanism fails for all
but unrealistically soft nuclear equations of state.

The delayed mechanism begins with a failed hydrodynamic explosion;
after about 0.01 seconds the shock wave stalls at a radius of 200-300
km.  It exists in a sort of equilibrium, gaining energy from matter
falling across the shock front, but losing energy to the heating of
that material.  However, after perhaps 0.5 seconds, the shock wave is
revived due to neutrino heating of the nucleon ``soup" left in the
wake of the shock.  This heating comes primarily from charged current
reactions off the nucleons in that nucleon gas; quasi-elastic
scattering also contributes.  This high entropy radiation-dominated
gas may reach two MeV in temperature.  The pressure exerted by this
gas helps to push the shock outward. It is important to note that
there are limits to how effective this neutrino energy transfer can
be: if matter is too far from the core, the coupling to neutrinos is
too weak to deposit significant energy.  If too close, the matter may
be at a temperature (or soon reach a temperature) where neutrino
emission cools the matter as fast or faster than neutrino absorption
heats it.  The term ``gain radius" is used to describe the region
where useful heating is done.

This subject is still controversial and unclear.  The problem is
numerically challenging, forcing modelers to handle the difficult
hydrodynamics of a shock wave; the complications of the nuclear
equation of state at densities not yet accessible to experiment;
modeling in two or three dimensions; handling the slow diffusion of
neutrinos; etc.  Not all of these aspects can be handled reasonably at
the same time, even with existing supercomputers.  Thus there is
considerable disagreement about whether we have any supernova model
that succeeds in ejecting the mantle.

However the explosion proceeds, there is agreement that 99\% of the 3
$\cdot 10^{53}$ ergs released in the collapse is radiated in neutrinos
of all flavors.  The time scale over which the trapped neutrinos leak
out of the protoneutron star is about three seconds.  Through most of
their migration out of the protoneutron star, the neutrinos are in
flavor equilibrium
\begin{equation}
 \mathrm{e.g.},~~ \nu_e + \bar{\nu}_e \leftrightarrow \nu_\mu + \bar{\nu}_\mu. 
\end{equation}
As a result, there is an approximate equipartition of energy among the
neutrino flavors.  After weak decoupling, the $\nu_e$s and
$\bar{\nu_e}$s remain in equilibrium with the matter for a longer
period than their heavy-flavor counterparts, due to the larger cross
sections for scattering off electrons and because of the
charge-current reactions
\begin{eqnarray}
 \nu_e + \mathrm{n} &&\leftrightarrow \mathrm{p} + e^- \nonumber \\ 
 \bar{\nu_e} + \mathrm{p} &&\leftrightarrow \mathrm{n} + e^+. 
\end{eqnarray}
Thus the heavy flavor neutrinos decouple from deeper within the star,
where temperatures are higher.  Typical calculations yield
\begin{equation}
 T_{\nu_\mu} \sim T_{\nu_\tau} \sim 8 \mathrm{MeV} ~~~~
 T_{\nu_e} \sim 3.5 \mathrm{MeV}~~~~T_{\bar{\nu_e}} \sim 4.5 \mathrm{MeV}. 
\end{equation}
The difference between the $\nu_e$ and $\bar{\nu_e}$ temperatures is a
result of the neutron richness of the matter, which enhances the rate
for charge-current reactions of the $\nu_e$s, thereby keeping them
coupled to the matter somewhat longer.  (This temperature hierarchy, particular
the difference between the heavy-flavor and electron neutrino
temperatures, is still a matter of some debate, as it is influenced
by both the explosion mechanism (e.g., by neutron fingers and
other types of mixing) and the detailed modeling of the
microphysics.  Some recent work~\cite{raffelt} argues for
considerably smaller temperature differences than those given
above.)

This temperature hierarchy is important because temperature 
inversions are a potential signature of oscillations.  A three-flavor
MSW level-crossing diagram is shown in fig. 15.  
Naively (that is, without considering neutrino-neutrino scattering
and other effects that could alter this picture) the crossings
corresponding to $\delta m^2_{solar}$ and $\delta m^2_{atmos}$
both occur well outside the neutrino sphere, that is, after the
neutrinos have decoupled and have fixed spectra characterized by the
temperatures given above.  (For example, $\delta m^2_{atmos}$
corresponds to a electron density typical of the base of the
carbon zone, prior to the explosion.)  
Thus a $\nu_e \leftrightarrow \nu_\tau$ 
oscillation would produce a distinctive $T \sim 8$ MeV spectrum of
$\nu_e$s.  Because neutrino-nucleus cross sections often grow 
as a high power of the neutrino energy (due both to phase space
and threshold effects), this will produce elevated $\nu_e$
event rates in many detectors.  Oscillations may also have an
effect on nucleosynthesis, such as the $\nu$-process we will
discuss below.

\begin{figure}[htb]
\psfig{bbllx=1.0cm,bblly=4.0cm,bburx=18cm,bbury=18.5cm,figure=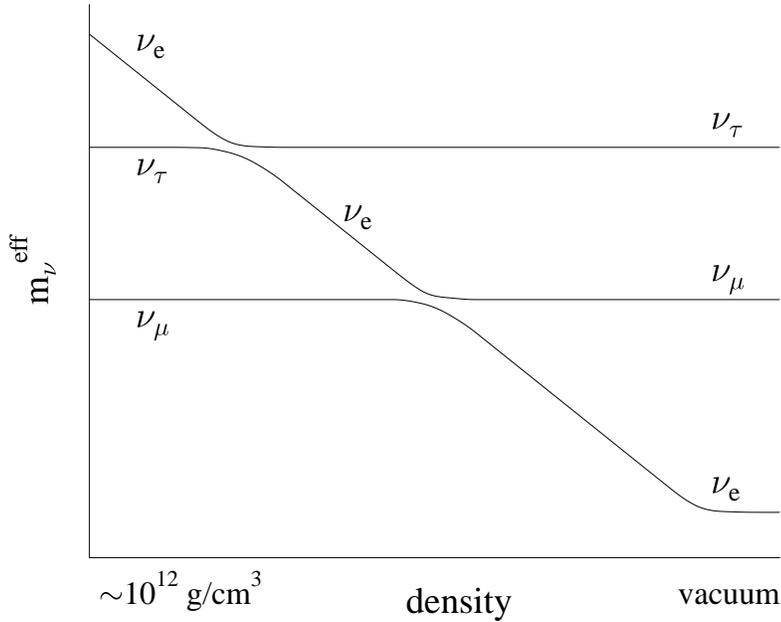,height=3.5in}
\caption{Three-flavor neutrino level-crossing diagram.  The 
illustrated $\nu_e \leftrightarrow \nu_\mu$ and 
$\nu_e \leftrightarrow \nu_\tau$ crossings should arise
from the solar and atmospheric $\delta m^2$s.  The diagram
illustrates that these crossing occur outside the neutrinosphere,
in regions where the neutrinos are fully decoupled from the
matter.}
\end{figure}
  
\subsection{The Neutrino Process~\protect\cite{nupro}}

Core-collapse supernovae are one of the major engines driving galactic
chemical evolution, producing and ejecting the metals that enrich our
galaxy.  The discussion of the previous section described the
hydrostatic evolution of a presupernova star in which large quantities
of the most abundant metals (C, O, Ne,...) are synthesized and later
ejected during the explosion.  During the passage of the shock wave
through the star's mantle, temperature of $\sim (1-3) \cdot 10^9$K and
are reached in the silicon, oxygen, and neon shells.  This shock wave
heating induces $(\gamma,\alpha) \leftrightarrow (\alpha,\gamma)$ and
related reactions that generate a mass flow toward highly bound
nuclei, resulting in the synthesis of iron peak elements as well as
less abundant odd-A species.  Rapid neutron-induced reactions are
thought to take place in the high-entropy atmosphere just above the
mass cut, producing about half of the heavy elements above A $\sim$
80.  Finally, the $\nu$-process
described below is responsible for the synthesis of rare species such
as $^{11}$B and $^{19}$F.  This process involves the weak response of
nuclei at momentum transfers where the allowed approximation is no
longer valid.  Thus we will use the $\nu$-process in this section to
illustrate some of the relevant nuclear physics.

One of the problems -- still controversial -- that may be connected
with the neutrino process is the origin of the light elements Be, B
and Li, elements which are not produced in sufficient amounts in the
big bang or in any of the stellar mechanisms we have discussed.  The
traditional explanation has been cosmic ray spallation interactions
with C, O, and N in the interstellar medium.  In this picture, cosmic
ray protons collide with C at relatively high energy, knocking the
nucleus apart.  So in the debris one can find nuclei like $^{10}$B,
$^{11}$B, and $^7$Li.

But there are some problems with this picture.  First of all, this is
an example of a secondary mechanism: the interstellar medium must be
enriched in the C, O, and N to provide the targets for these
reactions.  Thus cosmic ray spallation must become more effective as
the galaxy ages.  The abundance of boron, for example, would tend to
grow quadratically with metallicity, since the rate of production goes
linearly with metallicity.  But observations, especially recent
measurements with the HST, find a linear growth~\cite{timmes} in the
boron abundance.

A second problem is that the spectrum of cosmic ray protons peaks near
1 GeV, leading to roughly comparable production of the two isotopes
$^{10}$B and $^{11}$B.  That is, while it takes more energy to knock
two nucleons out of carbon than one, this difference is not
significant compared to typical cosmic ray energies.  More careful
studies lead to the expectation that the abundance ratio of $^{11}$B
to $^{10}$B might be $\sim$ 2.  In nature, it is greater than 4.

Fans of cosmic ray spallation have offered solutions to these
problems, e.g., similar reactions occurring in the atmospheres of
nebulae involving lower energy cosmic rays.  As this suggestion was
originally stimulated by the observation of nuclear $\gamma$ rays from
Orion, now retracted, some of the motivation for this scenario has
evaporated.  Here we focus on an alternative explanation, synthesis
via neutrino spallation.

Previously we described the allowed Gamow-Teller (spin-flip) and Fermi
weak interaction operators.  These are the appropriate operators when
one probes the nucleus at a wavelength -- that is, at a size scale --
where the nucleus responds like an elementary particle.  We can then
characterize its response by its macroscopic quantum numbers, the spin
and charge.  On the other hand, the nucleus is a composite object and,
therefore, if it is probed at shorter length scales, all kinds of
interesting radial excitations will result, analogous to the
vibrations of a drumhead.  For a reaction like neutrino scattering off
a nucleus, the full operator involves the additional factor
\begin{equation}
e^{i \vec{k} \cdot \vec{r}} \sim 1 + i \vec{k} \cdot \vec{r} 
\end{equation}
where the expression on the right is valid if the magnitude of
$\vec{k}$ is not too large.  Thus the full charge operator 
includes a ``first forbidden" term
\begin{equation}
 \sum_{i=1}^A \vec{r}_i \tau_3(i) 
\end{equation}
and similarly for the spin operator
\begin{equation}
 \sum_{i=1}^A [\vec{r}_i \otimes \vec{\sigma}(i)]_{J=0,1,2} \tau_3(i). 
\end{equation}
These operators generate collective radial excitations, leading to the
so-called ``giant resonance" excitations in nuclei.  The giant
resonances are typically at an excitation energy of 20-25 MeV in light
nuclei.  One important property is that these operators satisfy a sum
rule (Thomas-Reiche-Kuhn) of the form
\begin{equation}
 \sum_f | \langle f | \sum_{i=1}^A r(i) \tau_3(i) | i \rangle |^2
\sim {N Z \over A} \sim {A \over 4} 
\end{equation}
where the sum extends over a complete set of final nuclear states.
These first-forbidden operators tend to dominate the cross sections
for scattering the high energy supernova neutrinos ($\nu_{\mu}$s and
$\nu_\tau$s), with $E_\nu \sim$ 25 MeV, off light nuclei. From the sum
rule above, it follows that nuclear cross sections per target {\it
  nucleon} are roughly constant.

The E1 giant dipole mode described above is depicted qualitatively in
fig. 16a.  This description, which corresponds to an early model of
the giant resonance response by Goldhaber and Teller, involves the
harmonic oscillation of the proton and neutron fluids against one
another.  The restoring force for small displacements would be linear
in the displacement and dependent on the nuclear symmetry energy.
There is a natural extension of this model to weak interactions, where
axial excitations occur.  For example, one can envision a mode similar
to that of fig. 16a where the spin-up neutrons and spin-down protons
oscillate against spin-down neutrons and spin-up protons, the
spin-isospin mode of fig. 16b.  This mode is one that arises in a
simple SU(4) extension of the Goldhaber-Teller model, derived by
assuming that the nuclear force is spin and isospin independent, at
the same excitation energy as the E1 mode.  In full, the
Goldhaber-Teller model predicts a degenerate 15-dimensional
supermultiplet of giant resonances, each obeying sum rules analogous
to the TRK sum rule.  While more sophisticated descriptions of the
giant resonance region are available, of course, this crude picture is
qualitatively accurate.
  
\begin{figure}[htb]
\psfig{bbllx=0.3cm,bblly=2.8cm,bburx=13cm,bbury=11.5cm,figure=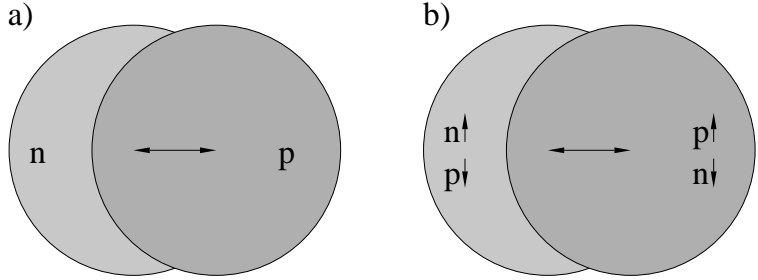,height=1.9in}
\caption{Schematic illustration of a) the E1 giant dipole mode
familiar from electromagnetic interactions and b) a spin-isospin
giant dipole mode associated with the first-forbidden weak
axial response.}
\end{figure}
  
This nuclear physics is important to the $\nu$-process.  The simplest
example of $\nu$-process nucleosynthesis involves the Ne shell in a
supernova.  Because of the first-forbidden contributions, the cross
section for inelastic neutrino scattering to the giant resonances in
Ne is $\sim 3 \cdot 10^{-41}$ cm$^2$/flavor for the more energetic
heavy-flavor neutrinos.  This reaction
\begin{equation}
 \nu + A \rightarrow \nu' + A^* 
\end{equation}
transfers an energy typical of giant resonances, $\sim$ 20 MeV.  A
supernova releases about 3 $\times 10^{53}$ ergs in neutrinos, which
converts to about $4 \times 10^{57}$ heavy flavor neutrinos.  The Ne
shell in a 20 M$_\odot$ star has at a radius $\sim$ 20,000 km.  Thus
the neutrino fluence through the Ne shell is
\begin{equation}
 \phi \sim { 4 \cdot 10^{57} \over 4 \pi (20,000 \mathrm{km})^2 }
\sim 10^{38}/\mathrm{cm}^2. 
\end{equation}
Thus folding the fluence and cross section, one concludes that
approximately 1/300th of the Ne nuclei interact.

This is quite interesting since the astrophysical origin of $^{19}$F
had not been understood.  The only stable isotope of fluorine,
$^{19}$F has an abundance
\begin{equation}
 {^{19}\mathrm{F} \over ^{20}\mathrm{Ne}} \sim {1 \over 3100}. 
\end{equation}
This leads to the conclusion that the fluorine found in a tube of
toothpaste was created by neutral current neutrino reactions deep
inside some ancient supernova.

The calculation of the final $^{19}$F/$^{20}$Ne ratio is
more complicated than the simple 1/300 ratio given above: \\
$\bullet$ When Ne is excited by $\sim$ 20 MeV through inelastic
neutrino scattering, it breaks up in two ways
\begin{eqnarray}
 ^{20}\mathrm{Ne}(\nu,\nu')^{20}\mathrm{Ne}^* 
&&\rightarrow ^{19}\mathrm{Ne} + n 
\rightarrow ^{19}\mathrm{F} + e^+ + \nu_e + n \nonumber \\
 ^{20}\mathrm{Ne}(\nu,\nu')^{20}\mathrm{Ne}^* 
&&\rightarrow ^{19}\mathrm{F}
+ p 
\end{eqnarray}
with the first reaction occurring half as frequently as the second.
As both channels lead to $^{19}$F, we have correctly estimated the
instantaneous abundance ratio in the Ne shell of
\begin{equation}
 {^{19}\mathrm{F} \over ^{20}\mathrm{Ne}} \sim {1 \over 300}. 
\end{equation}
$\bullet$ We must also address the issue of whether the produced
$^{19}$F survives.  In the first 10$^{-8}$ sec the co-produced neutrons
in the first reaction react via
\begin{equation}
^{15}\mathrm{O}(n,p)^{15}\mathrm{N}~~
^{19}\mathrm{Ne}(n,\alpha)^{16}\mathrm{O}~~
^{20}\mathrm{Ne}(n,\gamma)^{21}\mathrm{Ne}~~
^{19}\mathrm{Ne}(n,p)^{19}\mathrm{F} 
\end{equation}
with the result that about 70\% of the $^{19}$F produced via
spallation of neutrons is then immediate destroyed, primarily by the
$(n,\alpha)$ reaction above.  In the next $10^{-6}$ sec the co-produced
protons are also processed
\begin{equation}
 ^{15}\mathrm{N}(p,\alpha)^{12}\mathrm{C}~~
^{19}\mathrm{F}(p,\alpha)^{16}\mathrm{O}~~
^{23}\mathrm{Na}(p,\alpha)^{20}\mathrm{Ne} 
\end{equation}
with the latter two reactions competing as the primary proton poisons.
This makes an important prediction: stars with high Na abundances
should make more F, as the $^{23}$Na acts as a proton
poison to preserve the produced F.
$\bullet$ Finally, there is one other destruction mechanism, the
heating associated with the passage of the shock wave.  It turns out
the the F produced prior to shock wave passage can survive if it is in
the outside half of the Ne shell.  The reaction
\begin{equation}
 ^{19}\mathrm{F}(\gamma,\alpha)^{15}\mathrm{N} 
\end{equation}
destroys F for peak explosion temperatures exceeding $1.7 \cdot
10^9$K.  Such a temperature is produced at the inner edge of the Ne
shell by the shock wave heating, but not at the outer edge.

If all of this physics in handled is a careful network code that
includes the shock wave heating and F production both before and
after shock wave passage, the following are the results:
 \[ \begin{array}{cc} \underline{[^{19}\mathrm{F}/^{20}\mathrm{Ne}]/
[^{19}\mathrm{F}/^{20}\mathrm{Ne}]_\odot} & 
\underline{T_{\mathrm{heavy}~\nu} \mathrm{(MeV)}} \\
0.14 & 4 \\ 0.6 & 6 \\ 1.2 & 8 \\ 1.1 & 10 \\ 1.1 & 12 \end{array} \]
where the abundance ratio in the first column has been normalized to
the solar value. One sees that the attribution of F to the neutrino
process argues that the heavy flavor $\nu$ temperature must be greater
than 6 MeV, a result theory favors.  One also sees that F cannot be
overproduced by this mechanism: although the instantaneous production
of F continues to grow rapidly with the neutrino temperature, too much
F results in its destruction through the $(p,\alpha)$ reaction, given
the metalicity assumed in this calculation
(a solar abundance of the competing proton poison $^{23}$Na).  Indeed,
this illustrates an odd quirk: although in most cases the neutrino
process is a primary mechanism, one needs $^{23}$Na present to produce
significant F. Thus in this case the neutrino process is a secondary
mechanism.

While there are other significant neutrino process products ($^7$Li,
$^{138}$La, $^{180}$Ta, $^{15}$N ...), the most important product is
$^{11}$B, produced by spallation off carbon.  A calculation by Timmes
et al.\cite{timmes} found that the combination of the neutrino
process, cosmic ray spallation and big-bang nucleosynthesis together
can explain the evolution of the light elements.  The neutrino
process, which produces a great deal of $^{11}$B but relatively little
$^{10}$B, combines with the cosmic ray spallation mechanism to yield
the observed isotope ratio.  Again, one prediction of this picture is
that early stars should be $^{11}$B rich, as the neutrino process is
primary and operates early in our galaxy's history; the cosmic ray
production of $^{10}$B is more recent.  There is hope that HST studies
will soon be able to discriminate between $^{10}$B and $^{11}$B: as
yet this has not been done.

\subsection{The $r$-process~\protect\cite{qianr}}

Beyond the iron peak nuclear Coulomb barriers become so high that
charged particle reactions become ineffective, leaving neutron capture
as the mechanism responsible for producing the heaviest nuclei.  If
the neutron abundance is modest, this capture occurs in such a way
that each newly synthesized nucleus has the opportunity to $\beta$
decay, if it is energetically favorable to do so.  Thus weak
equilibrium is maintained within the nucleus, so that synthesis is
along the path of stable nuclei.  This is called the s- or
slow-process.  However a plot of the s-process in the (N,Z) plane
reveals that this path misses many stable, neutron-rich nuclei that
are known to exist in nature.  This suggests that another mechanism is
at work, too.  Furthermore, the abundance peaks found in nature near
masses A $\sim$ 130 and A $\sim$ 190, which mark the closed neutron
shells where neutron capture rates and $\beta$ decay rates are slower,
each split into two sub-peaks.  One set of sub-peaks corresponds to the
closed-neutron-shell numbers N $\sim$ 82 and N $\sim$ 126, and is
clearly associated with the s-process.  The other set is shifted to
smaller N, $\sim$ 76 and $\sim$ 116, respectively, and is suggestive
of a much more explosive neutron capture environment where neutron
capture can be rapid.
  
This second process is the r- or rapid-process, characterized by: \\
$\bullet$ The neutron capture is fast compared to $\beta$ decay
rates. \\
$\bullet$ The equilibrium maintained within a nucleus is established
by (n$,\gamma) \leftrightarrow (\gamma,$n): neutron capture fills up
the available bound levels in the nucleus until this equilibrium sets
in.  The new Fermi level
depends on the temperature and the relative n$/\gamma$ abundance.\\
$\bullet$ The nucleosynthesis rate is thus controlled by the $\beta$
decay rate: each $\beta^-$ capture converting n $\rightarrow$ p opens
up a hole in the neutron Fermi sea, allowing another neutron
to be captured. \\
$\bullet$ The nucleosynthesis path is along exotic, neutron-rich
nuclei that would be highly unstable under normal laboratory
conditions. \\
$\bullet$ As the nucleosynthesis rate is controlled by the $\beta$
decay, mass will build up at nuclei where the $\beta$ decay rates are
slow.  It follows, if the neutron flux is reasonable steady over time
so that equilibrated mass flow is reached, that the resulting
abundances should be inversely proportional to these $\beta$ decay
rates.
  
Let's first explore the (n$,\gamma) \leftrightarrow (\gamma,$n)
equilibrium condition, which requires that the rate for (n$,\gamma)$
balances that for $(\gamma,$n) for an average nucleus.  So consider
the formation cross section
\begin{equation}
 A + \mathrm{n} \rightarrow (A+1) + \gamma . 
\end{equation}
This is an exothermic reaction, as the neutron drops into the nuclear
well.  Our averaged cross section, assuming a resonant reaction (the
level density is high in heavy nuclei) is
\begin{equation}
\langle \sigma v \rangle_{(\mathrm{n},\gamma)} = 
\left( {2 \pi \over \mu kT} \right)^{3/2} {\Gamma_{\mathrm{n}} \Gamma_\gamma
\over \Gamma} e^{-E/KT} 
\end{equation}
where E $\sim$ 0 is the resonance energy,
and the $\Gamma$s are the indicated partial and total widths.
Thus the rate per unit volume is
\begin{equation}
r_{(\mathrm{n},\gamma)} \sim N_{\mathrm{n}} N_A \left( {2 \pi \over \mu kT} \right)^{3/2}
{\Gamma_{\mathrm{n}} \Gamma_\gamma \over \Gamma}
\end{equation}
where $N_{\mathrm{n}}$ and $N_A$ are the neutron and nuclear number densities
and $\mu$ the reduced mass.
This has to be compared to the $(\gamma,$n) rate. 

The $(\gamma,$n) reaction requires the photon number density in
the gas.  This is given by the Bose-Einstein distribution
\begin{equation}
N(\epsilon) = {8 \pi \over c^3 h^3} {\epsilon^2 d \epsilon
\over e^{\epsilon/kT} -1} .
\end{equation}
The high-energy tail of the normalized distribution can thus
be written
\begin{equation}
 \sim {1 \over N_\gamma \pi^2} \epsilon^2 e^{-\epsilon/kT} d \epsilon 
\end{equation}
where in the last expression we have set $\hbar = c = 1$. 

Now we need the resonant cross section in the $(\gamma,$n) direction.
For photons the wave number is proportional to the energy, so
\begin{equation}
\sigma_{(\gamma,\mathrm{n})} = {\pi \over \epsilon^2}
{\Gamma_\gamma \Gamma_{\mathrm{n}} \over (\epsilon-E_r)^2 + (\Gamma/2)^2 } .
\end{equation}
As the velocity is c =1,
\begin{equation}
\langle \sigma v \rangle = {1 \over \pi^2 N_\gamma}
\int_0^\infty \epsilon^2 e^{-\epsilon/kT} d \epsilon
{\pi \over \epsilon^2} {\Gamma_\gamma \Gamma_{\mathrm{n}} \over 
(\epsilon-E_r)^2 +(\Gamma/2)^2} . 
\end{equation}
We evaluate this in the usual way for a sharp resonance, remembering
that the energy integral over just the denominator above (the sharply
varying part) is $2 \pi/ \Gamma$
\begin{equation}
 \sim {\Gamma_\gamma \Gamma_{\mathrm{n}} \over N_\gamma} e^{-E_r/kT}
{2 \over \Gamma} . 
\end{equation}
So that the rate becomes
\begin{equation}
r_{(\gamma,\mathrm{n})} \sim 2 N_{A+1} {\Gamma_\gamma \Gamma_{\mathrm{n}}
\over \Gamma} e^{-E_r/kT} .  
\end{equation}
Equating the (n$,\gamma)$ and $(\gamma,$n) rates and taking $N_A \sim
N_{A-1}$ then yields
\begin{equation}
N_{\mathrm{n}} \sim {2 \over (\hbar c)^3} \left( {\mu c^2 kT \over
2 \pi} \right)^{3/2} e^{-E_r/kT} 
\end{equation}
where the $\hbar$s and $c$s have been properly inserted to give the
right dimensions.  Now $E_r$ is essentially the binding energy.  So
plugging in the conditions $N_{\mathrm{n}} \sim 3 \times 10^{23}$/cm$^3$ and $T_9
\sim 1$, we find that the binding energy is $\sim$ 2.4 MeV.  Thus
neutrons are bound by about 30 times $kT$, a value that is still small
compared to a typical binding of 8 MeV for a normal nucleus.  (In this
calculation the neutron reduced mass is calculated by assuming a
nuclear target with A=150.)

The above calculation fails to count spin states for the photons and
nuclei and is thus not quite correct.  But it makes the essential
point: the $r$-process involves very exotic species largely unstudied in
any terrestrial laboratory.  It is good to bear this in mind, as in
the following section we will discuss the responses of such nuclei to
neutrinos.  Such responses thus depend on the ability of theory to
extrapolate responses from known nuclei to those quite unfamiliar.

The path of the $r$-process is along neutron-rich nuclei, where the
neutron Fermi sea is just $\sim$ (2-3) MeV away from the neutron drip
line (where no more bound neutron levels exist).  After the $r$-process
finishes (the neutron exposure ends) the nuclei decay back to the
valley of stability by $\beta$ decay.  This can involve some neutron
spallation ($\beta$-delayed neutrons) that shift the mass number A to
a lower value.  But it certainly involves conversion of neutrons into
protons, and that shifts the $r$-process peaks at N $\sim$ 82 and 126 to
a lower N, off course.  This effect is clearly seen in the abundance
distribution: the $r$-process peaks are shifted to lower N relative to
the s-process peaks.  This is the origin of the second set of
``sub-peaks" mentioned at the start of the section.

It is believed that the $r$-process can proceed to very heavy nuclei (A
$\sim$ 270) where it is finally ended by $\beta$-delayed and n-induced
fission, which feeds matter back into the process at an A $\sim$
A$_{max}$/2.  Thus there may be important cycling effects in the upper
half of the $r$-process distribution.
  
What is the site(s) of the $r$-process?  This has been debated 
many years and still remains a controversial subject:\\
$\bullet$ The $r$-process requires exceptionally explosive conditions 
\begin{center}
$\rho$(n) $\sim 10^{20}$ cm$^{-3}$~~~T $\sim 10^9$K~~~t $\sim$ 1s.
\end{center}
$\bullet$ Both have been primary and secondary sites have been proposed.  Primary sites
are those not requiring preexisting metals.  Secondary sites are those
where the neutron capture occurs
on preexisting s-process seeds.\\
$\bullet$ Suggested primary sites include the the neutronized
atmosphere above the proto-neutron star in a Type II supernova,
neutron-rich jets produced in supernova explosions or in neutron star
mergers, inhomogeneous big
bangs, etc. \\
$\bullet$ Secondary sites, where $\rho$(n) can be lower for successful
synthesis, include the He and C zones in Type II supernovae, the red
giant He flash, etc.

The balance of evidence favors a primary site, so one requiring
no pre-enrichment of heavy s-process metals.  Among the evidence: \\
  
\noindent
1) Keck and HST studies of very-metal-poor halo stars: The most important
evidence are the recent measurements of Cowan, Sneden et
al.~\cite{sneden} of very metal-poor stars ([Fe/H] $\sim$ -1.7 to
-3.12) where an $r$-process distribution very much like that of our sun
has been seen for Z $\gsim$ 56.  Furthermore, in these stars the iron
content is variable.  This suggests that the ``time resolution"
inherent in these old stars is short compared to galactic mixing times
(otherwise Fe would be more constant).  The conclusion is that the
$r$-process material in these stars is most likely from one or a few
local supernovae.  The fact that the distributions match the solar
$r$-process (at least above charge 56) strongly suggests that there is
some kind of unique site for the high-Z portion of the $r$-process: the solar $r$-process
distribution did not come from averaging over many different kinds of
$r$-process events.  Clearly the fact that these old stars are enriched
in $r$-process metals also strongly argues for a primary process: the
$r$-process works quite well in an
environment where there are few initial s-process metals.
(The situation for elements below Z=56 is somewhat different.  While
an adequate discussion would take us beyond the limits of
these lectures,
interested reader are directed to Ref. [42].)

\noindent
2) There are also fairly good theoretical arguments that a primary
$r$-process occurring in a core-collapse supernova might be
viable~\cite{hotbub}.  First, galactic chemical evolution studies
indicate that the growth of $r$-process elements in the galaxy is
consistent with low-mass Type II supernovae in rate and distribution.
More convincing is the fact that modelers have shown that the
conditions needed for an $r$-process (very high neutron densities,
temperatures of 1-3 billion degrees) might be realized in a supernova.
The site is the last material expelled from the supernova, the matter
just above the mass cut.  When this material is blown off the star
initially, it is a very hot neutron-rich, radiation-dominated gas
containing neutrons and protons, but an excess of the neutrons.  As it
expands off the star and cools, the material first goes through a
freeze-out to $\alpha$ particles, a step that essentially locks up all
the protons in this way.  Then the $\alpha$s interact through
reactions like
\begin{eqnarray}
 \alpha + \alpha +\alpha &&\rightarrow {}^{12}\mathrm{C}  \nonumber \\
 \alpha + \alpha + n &&\rightarrow {}^9\mathrm{Be} \nonumber
\end{eqnarray}
to start forming heavier nuclei.  Note, unlike the big bang,
that the density is high enough to allow such three-body 
interactions to bridge the mass gaps at A = 5,8.  The
$\alpha$ capture continues up to heavy nuclei,
to A $\sim$ 80, in the network calculations.  
The result is a small number of ``seed" nuclei,
a large number of $\alpha$s, and excess neutrons.  These 
neutrons preferentially capture on the heavy seeds to
produce an $r$-process.  Of course, what is necessary is to
have $\sim$ 100 excess neutrons per seed in order to 
successfully synthesize heavy mass nuclei.  Some of the
modelers find conditions where this almost happens. 
  
There are some very nice aspects of this site: the amount of matter
ejected is about 10$^{-5} - 10^{-6}$ solar masses, which is just about
what is needed over the lifetime of the galaxy to give the integrated
$r$-process metals we see, taking a reasonable supernova rate.  But
there are also a few problems, especially the fact that with
calculated entropies in the nucleon soup above the proto-neutron star,
neutron fractions appear to be too low to produce a successful A
$\sim$ 190 peak.  There is some interesting recent work invoking
neutrino oscillations~\cite{gail1} to cure this problem: charge current reactions
on free protons and neutrons determine the n/p ratio in the gas.
This is discussed briefly in the next section.

The nuclear physics of the $r$-process tells us that the synthesis
occurs when the nucleon soup is in the temperature range of (3-1)
$\cdot 10^9$K, which, in the hot bubble $r$-process described above,
corresponds to a freeze-out radius of (600-1000) km and a time $\sim$ 10
seconds after core collapse.  The neutrino fluence after freeze-out
(when the temperature has dropped below 10$^9$K and the $r$-process
stops) is then $\sim$ (0.045-0.015) $\cdot 10^{51}$ ergs/(100km).
Thus, after completion of the $r$-process, the newly synthesized
material experiences an intense flux of neutrinos.  This brings up the
question of whether the neutrino flux could have any effect on the
$r$-process.

\subsection{Neutrinos and the $r$-process~\protect\cite{qian}}

Before describing the exotic effects of neutrino oscillations on
the supernovae and nucleosynthesis, we will examine standard-model
effects that are nevertheless quite interesting.  The nuclear physics
of this section -- neutrino-induced neutron spallation reactions -- is
also relevant to recently proposed supernova neutrino observatories
such as OMNIS and LAND.  Comparing to our earlier discussion of carbon- and neon-zone
synthesis by the $\nu$-process, it is apparent that neutrino effects could
be much larger in the hot bubble $r$-process: the synthesis occurs {\it
  much} closer to the star than our Ne radius of 20,000 km: estimates
are 600-1000 km.  The $r$-process is completed in about 10 seconds (when
the temperature drops to about one billion degrees), but the neutrino
flux is still significant as the $r$-process freezes out.  The net
result is that the ``post-processing" neutrino fluence - the fluence
that can alter the nuclear distribution after the $r$-process is
completed - is about 100 times larger than that responsible for
fluorine production in the Ne zone.  Recalling that 1/300 of the
nuclei in the Ne zone interacted with neutrinos, and remembering that
the relevant neutrino-nucleus cross sections scale as A, one quickly
sees that the probability of a $r$-process nucleus interacting with the
neutrino flux is approximately unity.

Because the hydrodynamic conditions of the $r$-process are highly
uncertain, one way to attack this problem is to work backward in time.
We know the final $r$-process distribution (what nature gives us) and we
can calculate neutrino-nucleus interactions relatively well.  Thus
from the observed $r$-process distribution (including neutrino
post-processing) we can work backward to find out what the $r$-process
distribution looked like at the point of freeze-out.  In Figs. 17 and
18, the ``real" $r$-process distribution - that produced at freeze-out -
is given by the dashed lines, while the solid lines show the effects
of the neutrino post-processing for a particular choice of fluence.
The nuclear physics input into these calculations is precisely that
previously described: GT and first-forbidden cross sections, with the
responses centered at excitation energies consistent with those found
in ordinary, stable nuclei, taking into account the observed
dependence on $|N-Z|$.

\begin{figure}[htb]
\psfig{bbllx=-2.0cm,bblly=4.5cm,bburx=18cm,bbury=23.0cm,figure=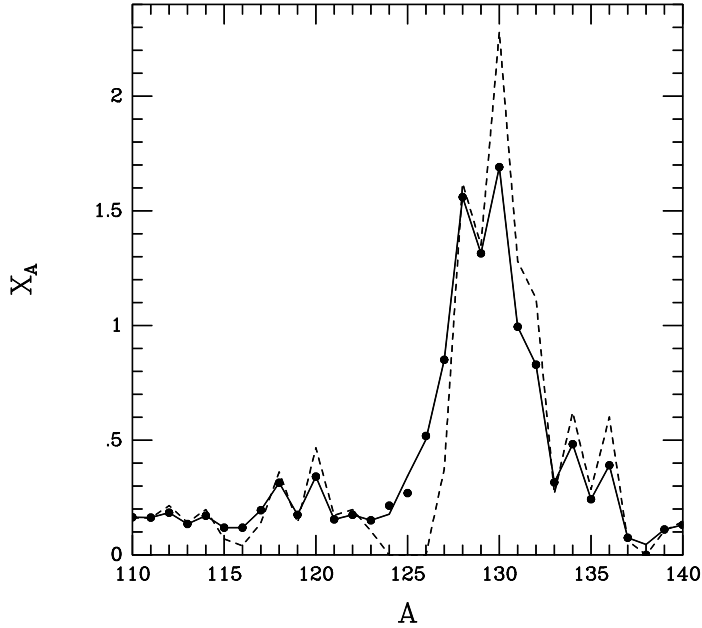,height=3.5in}
\caption{Comparison of the $r$-process distribution that would 
  result from the freeze-out abundances near the A $\sim$ 130 mass peak
  (dashed line) to that where the effects of neutrino post-processing
  have been include (solid line).  The fluence has been fixed by
  assuming that the A = 124-126 abundances are entirely due to the
  $\nu$-process.}
\end{figure}
  
One important aspect of the figures is that the mass shift is
significant.  This has to do with the fact that a 20 MeV excitation of
a neutron-rich nucleus allows multiple neutrons ( $\sim$ 5) to be
emitted.  (Remember we found that the binding energy of the last
neutron in an $r$-process neutron-rich nuclei was about 2-3 MeV under
typical $r$-process conditions.)  The second thing to notice is that the
relative contribution of the neutrino process is particularly
important in the ``valleys" beneath the mass peaks: the reason is that
the parents on the mass peak are abundant, and the valley daughters
rare.  In fact, it follows from this that the neutrino process effects
can be dominant for precisely seven isotopes (Te, Re, etc.) lying in
these valleys.  Furthermore if an appropriate neutrino fluence is
picked, these isotope abundances are produced perfectly (given the
abundance errors).  The fluences are
\begin{eqnarray}
     \mathrm{N} &=& 82~ \mathrm{peak}~~~~~0.031 \cdot 10^{51} 
\mathrm{ergs/(100km)^2/flavor} \nonumber \\
     \mathrm{N} &=& 126~ \mathrm{peak}~~~~0.015 \cdot 10^{51} 
\mathrm{ergs/(100km)^2/flavor}, \nonumber
\end{eqnarray}
values in fine agreement with those that would be found in a hot
bubble $r$-process.  So this is circumstantial but significant evidence
that the material near the mass cut of a Type II supernova is the site
of the $r$-process: there is a neutrino fingerprint.

\begin{figure}[htb]
\psfig{bbllx=-2.0cm,bblly=4.5cm,bburx=18cm,bbury=23.0cm,figure=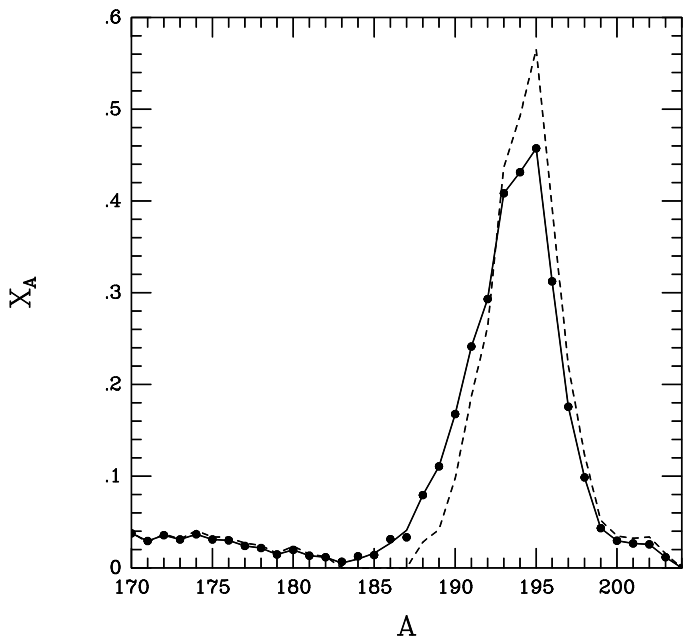,height=3.5in}
\caption{As in fig. 17, but for the A $\sim$ 195 mass peak.
The A = 183-187 abundances are entirely attributed to the 
$\nu$-process.}
\end{figure}

\section{Neutrino Oscillation Constraints from the $r$-process}
Many aspects of supernova physics could be altered by neutrino
oscillations.  One satisfying consequence of the recent solar
and atmospheric neutrino oscillation discoveries is that the 
derived parameters, the mass differences and mixing angles,
have begun to narrow some of ``parameter space'' for such
supernova oscillation effects.

Earlier it was noted that understanding neutrino transport 
is an essential but highly nontrivial part of the supernova
mechanism.  The addition of neutrino oscillations adds another
layer of complication, and thus opportunity for surprises.  To
set the stage for this discussion, we recall that
the density at the neutrinosphere is $\sim 10^{12}$g cm$^{-3}$
and the density at the position of the stalled shock is \cite{Mayle}
$\sim 2 \times 10^7$ g cm$^{-3}$. Writing the MSW resonance density in
appropriate units
\begin{equation}
\rho_{\rm res} = 1.31 \times 10^6 \left( {\delta m^2 \over {\rm eV}^2}
\right)  \left( { {\rm 10 MeV} \over E_{\nu} } \right) \left( {0.5 \over
{\rm Y}_e}  \right) \cos 2\theta~~ {\rm g} \> {\rm cm}^{-3},
\end{equation}
one sees, for $E_{\nu} \sim 10$ MeV, 
that the range of $\delta m^2$s producing crossings between the
neutrinosphere and the stalled shock wave is $\sim 1-10^4$ eV$^2$.
Some time ago Fuller {\it et al.}~\cite{Mayle} pointed out that $\nu_e-\nu_{heavy}$
oscillations in this region could substantial increase the 
explosion energy, as hot $\nu_e$s would couple more strongly to
the matter because of the increase in the $\nu_e + n$ cross
section (which goes as $E_\nu^2$, roughly).  Unfortunately the necessary range
of $\delta m^2$ includes neither the atmospheric nor solar
values.  Thus at least in the simplest scenario of three light
neutrinos, the oscillations we know about from experiment may
not affect this aspect of the explosion mechanism.  (However
there are many scenarios -- a fourth neutrino motivated by 
LSND \cite{lsnd}, CPT violation, etc -- where this conclusion 
would need to be reexamined.  Even in the simplest three-flavor
scenario discussed above, one should consider complications
that might move the MSW resonance position, such as the 
contributions of $\nu-\nu$ scattering to the MSW potential.)

The potential for neutrino oscillation to alter the $r$-process
(if core-collapse supernovae are indeed the site) was also 
recognized some time ago.  The $r$-process requires a 
neutron-rich environment: the ratio of
electrons to baryons, $Y_e$, should be less than one half. $Y_e$ in
the nucleosynthesis region is given approximately \cite{fuller} by 
\begin{equation}
Y_e \simeq {1 \over 1+ \lambda_{{\overline \nu}_e p} / \lambda_{ \nu_e
n}}  \simeq {1 \over 1 + T_{{\overline \nu}_e} / T_{ \nu_e}}, 
\end{equation}
where $\lambda_{ \nu_e n}$, etc. are the capture rates. Hence if
$T_{{\overline \nu}_e} > T_{\nu_e}$, then the medium is
neutron rich.  But oscillations of the type $\nu_e \leftrightarrow
\nu_{heavy}$ could invert this hierarchy, thus destroying 
a necessary condition for the $r$-process.  In the simplest 
scenario of three light neutrinos and $\delta m^2$ consistent
with the solar and atmospheric $\nu$ results, this catastrophe
is avoided for the hot-bubble $r$-process because the crossings
occur outside the region of interest.  (However, the same
caveats noted above apply.)

If the supernova is an $r$-process site it is also desirable to have
a neutron to seed-nucleus ratio ${\ 
  \lower-1.2pt\vbox{\hbox{\rlap{$>$}\lower5pt\vbox{\hbox{$\sim$}}}}\ }
100$ in order that the heavier $r$-process species ({\it i.e.}, those
in the $A=195$ peak) can be produced. This ratio is basically
determined by three quantities: i) the expansion
rate; ii) the electron fraction $Y_e$; and iii) the entropy per
baryon. Though different calculations \cite{hotbub,r1} disagree on the
value of the entropy in the neutrino-driven wind during the $r$-process
nucleosynthesis, several models can produce values of these three
parameters adequate for a successful $r$-process -- that is, until   
certain $\nu$ reactions are considered.  These $\nu$ reactions,
acting during or immediately after freezeout, allow too many
seed nuclei to form, thus producing an unfavorable neutron/seed
ratio.  This prevents the $r$-process from synthesizing the
heaviest nuclei, those in the $A \sim$ 190 peak.  The worst of
these $\nu$ reactions is called the alpha effect.

The alpha effect~\cite{meyeralpha} occurs at the epoch of alpha particle formation. As
the temperature drops, essentially all of the protons and most of the
neutrons in the ejecta lock themselves into tightly bound alpha particles.
As the matter was initially neutron rich, free neutrons remain and
$Y_e$ is below 0.5.  The alpha effect
pushes the electron fraction higher, towards $Y_e=0.5$. The increase in $Y_e$
comes about because $\nu_e$s capture on neutrons, with the produced
protons capturing more neutrons to produce $\alpha$s, thereby
reducing the number of free neutrons available for the
$r$-process \cite{alpha}. This effect has been shown to be the biggest
impediment to achieving an acceptable $r$-process yield \cite{MMF}. 
Matter-enhanced
neutrino oscillations between electron neutrinos and other active
species worsen this problem, tending to increase electron neutrino
energies and thus the average charge current cross section.
On the other hand, active-sterile oscillations which
diminish the flux of electron neutrinos outside some radius can
reduce the alpha effect, and thus preserve the $r$-process.  This
scenario has been discussed extensively because  
sterile neutrinos may also account for the LSND
results.

There is an important constraint on this ``solution,'' however.
In models where the $r$-process material is blown off the
protoneutron star by a neutrino wind, a
large flux of electron neutrinos is essential to overcome the 
binding effects of gravity.  As nucleons are
gravitationally bound by about $\sim100\,{\rm MeV}$ near the surface
of the protoneutron star while each neutrino has an energy $\sim
10$ MeV, $\sim$ 10 charge-current $\nu$-nucleon interactions are needed to 
eject the nucleon to infinity.  This in turn requires that the
active-sterile oscillation occur at a relatively large radius,
so that strong wind effects at smaller $r$ are unaffected.
It proves possible to arrange the necessary condition through
active-sterile $\nu_e
\rightleftharpoons \nu_s$ and $\bar\nu_e \rightleftharpoons
\bar{\nu}_s$ channels. In such a scheme \cite{gail1} the lightest
  sterile neutrino would be heavier than the $\nu_e$ and split from it
  by a vacuum mass-squared difference of 3 eV$^2 {\ 
    \lower-1.2pt\vbox{\hbox{\rlap{$<$}\lower5pt\vbox{\hbox{$\sim$}}}}\ 
    } \delta m^2_{es} {\ 
    \lower-1.2pt\vbox{\hbox{\rlap{$<$}\lower5pt\vbox{\hbox{$\sim$}}}}\ 
    }$ 70 eV$^2$ with vacuum mixing angle $\sin^2 2\theta_{es} >
  10^{-4}$.
Whether this solution is {\it necessary}, though, is quite 
another matter.  It has been argued that very fast expansion
rates could circumvent the alpha effect by minimizing the 
$\nu$ fluence experienced by the matter.  This underscores how 
difficult it is to assess our current understanding of the
$r$-process when our theoretical models of supernovae seem to
be less successful than Nature in producing explosions.
(This reminds one that the foundation for our discoveries in
solar neutrino physics was a reliable SSM!)

Finally, regardless of where the crossings occur within a 
supernova, the supernovae neutrinos that SNO, SuperKamiokande, and
other detectors record will have been altered by oscillations,
including potentially oscillation channels we have not yet
probed elsewhere (e.g., $\theta_{13}$).  The extent to which we
can exploit neutrinos as a probe of supernova physics 
will depend on how well we succeed with current efforts to
determine the entries in the mixing matrix.

This work was supported in part by the US Department of Energy 
under grants DE-FG03-00ER41132 and DE-FG02-01ER41187.  Portions of these lectures were based
on summer school notes prepared in collaboration
with A. Baha Balantekin, whom I thank.

\end{document}